\documentclass[12pt]{article}
\usepackage{latexsym}
 \csname
@addtoreset\endcsname{equation}{section}

\def\bec{\begin{center}}
\def\ec{\end{center}}
\def\a{\alpha} \def\ad{\dot{\a}} \def\hA{{\widehat A}}

\def\b{\beta}  \def\bd{\dot{\b}} 
\def\c{\gamma} \def\cd{\dot{\c}}
\def\C{\Gamma}
\def\d{\delta} \def\dd{\dot{\d}}

\def\e{\epsilon}

\def\F{\Phi}

\def\k{\kappa}
\def\l{\lambda}
\def\L{\Lambda}
\def\m{\mu}
\def\n{\nu}
\def\r{\rho}
\def\s{\sigma}

\def\t{\tau}

\def\x{\xi}
\def\y{\eta}

\def\O{\Omega}
\def\o{\omega}

\def\hF{\hat{\F}}

\def\yb{{\bar y}}
\def\zb{{\bar z}}

\def\ra{\rightarrow}

\let\la=\label

\def\nn{\nonumber}
\newcommand{\eq}[1]{(\ref{#1})}

\newcommand{\w}[1]{\\[0.#1cm]}
\def\be{\begin{equation}}
\def\ee{\end{equation}}
\def\bea{\begin{eqnarray}}
\def\eea{\end{eqnarray}}
\def\ba{\begin{array}}
\def\ea{\end{array}}

\def\mx#1#2#3#4{\left#1\begin{array}{#2} #3 \end{array}\right#4}

\def\ft#1#2{{\textstyle{{\scriptstyle #1}
\over {\scriptstyle #2}}}}

\def\scs#1{\section{\sc \large #1}}
\def\scss#1{\subsection{\sc  #1}}

\def\ad{\dot\alpha}
\def\bd{\dot\beta}
\def\sb{\bar\sigma}

\begin{document}
\begin{flushright}

MIFP-05-18\\
UUITP-12/05 \\
hep-th/0508158 \vskip 8pt


\end{flushright}

\vspace{10pt}

\begin{center}


{\Large\sc An Exact Solution of 4D Higher-Spin Gauge Theory}


\vspace{20pt}
{\sc E. Sezgin$^{1}$ and P. Sundell$^{2}$}\\[15pt]

{${}^1$\it\small George P. and Cynthia W. Mitchell Institute for
Fundamental
 Physics\\
Texas A\&M University\\ College Station, TX 77843-4242, USA}

{${}^2$\it\small Department for Theoretical Physics\\ Uppsala
University\\ Box 803, 751 08 Uppsala, SWEDEN} \vspace{5pt}


\vspace{15pt} {\sc\large Abstract}\end{center}

We give a one-parameter family of exact solutions to 4D
higher-spin gauge theory invariant under a deformed higher-spin
extension of $SO(3,1)$ and parameterized by a zero-form invariant.
All higher-spin gauge fields vanish, while the metric interpolates
between two asymptotically $AdS_4$ regions via two $dS_3$-foliated
domain walls and two $H_3$-foliated Robertson-Walker spacetimes --
one in the future and one in the past -- with the scalar field
playing the role of foliation parameter. All Weyl tensors vanish,
including that of spin two. We furthermore discuss methods for
constructing solutions, including deformation of solutions to pure
AdS gravity, the gauge-function approach, the perturbative
treatment of (pseudo-)singular initial data describing isometric
or otherwise projected solutions, and zero-form invariants.

\setcounter{page}{1}

\pagebreak

\tableofcontents


\scs{Introduction and Summary}\label{sec:in}


Full higher-spin gauge-field equations have been known in $D=4$
for quite some time, essentially since the early work of Vasiliev
\cite{vasiliev} (see \cite{Vas:star} for a review), and their
generalizations to higher dimensions have been started to be
understood more recently in
\cite{Vas:2003,Sagnotti:2005ns,Bekaert:2005vh}. These equations
are generalizations of pure AdS gravity, in which the metric is
accompanied by an infinite tower of higher-spin fields \emph{and}
special sets of lower-spin fields, always containing at least one
real scalar. In the minimal setting, the physical fields are
symmetric doubly traceless Lorentz tensors $\phi_{\mu_1\dots
\mu_s}$ of rank $s=0,2,4,\dots$. The generally covariant form of
the equations \cite{Us:analysis} (with non-linearly treated metric
$ds^2=dx^\mu dx^\nu g_{\m\n}$ accommodating the rank-two tensor)
is highly non-local with higher-derivative corrections normalized
by the AdS mass-parameter \cite{vasiliev,Vas:star}. The expansion
around the AdS vacuum yields, however, a tachyon and ghost-free
spectrum of massless particles governed by Fronsdal's equations,
with all non-localities occurring via interactions. In other
words, the classical higher-spin gauge theory is weakly coupled in
field amplitudes and strongly coupled in wave-numbers. This state
of affairs -- which refers to a classical theory although it
appears reminiscent to that of a quantum-effective theory -- blurs
many basic field-theoretic concepts, even at the level of the
lower-spin self-couplings, which, for example, no longer exhibit
any well-defined notion of microscopic scalar-field potential
\cite{Sezgin:2003pt} or local stress-energy tensor
\cite{Kristiansson:2003xx}. The non-localities are, of course, not
put in by hand, nor by integrating out microscopic fields.
Instead, the complete form of the interactions are governed --
somewhat in the spirit of classical string field theory -- by a
manifestly higher-spin covariant master-field formulation
\cite{vasiliev,Vas:star} (without ambiguities in case manifest
Lorentz invariance \cite{Vas:star} and parity invariance
\cite{Vas:more,Sezgin:2003pt} is required) based on simple
non-commutative twistor variables \cite{vasiliev} (see also
\cite{5d,5dn4,7d,Vasiliev:2004cm,Cederwall:2004cf,Bars:2005ze} for
higher-dimensional generalizations albeit at the free level)
related to deformation quantization of singletons
\cite{Bars:2001uy,Johan}.

The master-field formulation, which is presented in Section
\ref{sec:2}, is amenable to finding classical solutions, while the
non-localities make these difficult to interpret using traditional
field-theoretic and geometric tools. Clearly, the finding of exact
solutions is therefore highly desirable, not only for the
important role that they may play in uncovering novel physical
phenomena, but also for developing the interpretational language
as well as shedding new light on the origin of the higher-spin
field equations themselves in tensionless limits of String Theory
\cite{sundborg, sundell2,polyakovklebanov,Sagnotti:2003qa,
Beisert:2004di,Johan}.

It is primarily with the above motivations in mind that we have
sought and found an exact solution beyond AdS spacetime. In doing
so, we have greatly benefitted from the work of
\cite{Prokushkin:1998bq,Bolotin:1999fa,Vas:star} on how to use the
master-field formulation to construct solutions in ordinary
spacetime. Exact, massively deformed, 3D vacua are constructed in
\cite{Prokushkin:1998bq} using techniques for deforming twistor
oscillators, and plane-wave solutions for the free 4D field
equations are provided in \cite{Bolotin:1999fa} using a
gauge-function approach, in its turn related to a more general
method described in \cite{Prokushkin:1998bq,Vas:star} pertinent to
finding exact solutions in a systematic manner by means of
integrating flows. We shall relate further to these works when we
discuss methods for finding solutions in Section \ref{sec:3}.

The solution, which is presented in detail in Section \ref{sec:4},
is constructed by exploiting the simplifications taking place at
the full master-field level by imposing $SO(3,1)$-invariance
(without having to resort to weak-field expansion). As a result,
it is globally well-defined on a manifold with the same topology
as $AdS_4$, covered by two charts with local stereographic
coordinates $x^a$ and $\widetilde x^a$ (restricted by $\l^2
x^2,\l^2\widetilde x^2<1$ and related by $\widetilde x^a
=-x^a/(\lambda^2 x^2)$ in the overlap region $\lambda^2
x^2,\lambda^2\widetilde x^2<0$). In the Type A model (see Section
\ref{sec:mfe}), the solution is given locally in the $x^a$-chart
by
\be \phi(x)\ =\ {\nu\over b_1}(1-\l^2 x^2)\ ,\qquad ds^2\ =\
{4\O^2 (d(g_1 x))^2\over (1-\l^2g_1^2 x^2)^2}\ ,\qquad
\o_{\m}{}^{ab}\ =\ f \o_{(0)\m}^{ab}\ ,\label{intro:sol}\ee\be
\phi_{\m_1\dots \m_s}\ =\ 0\ ,\quad\mbox{for $s=4,6,\dots$}\ ,\ee
where the scale factors $\O(x^2;\nu)$, $g_1(x^2;\nu)$ and
$f(x^2;\nu)$ are given explicitly in \eq{f} and \eq{scale}, and
$\lambda$ is the inverse radius of the $AdS_4$ vacuum at $\nu=0$,
where $\O(x^2;0)=f(x^2;0)= g_1(x^2;0)=1$.

Remarkably, the higher-spin fields vanish, notwithstanding the
fact that lower spins in general source higher spins, so that the
field equations can only be truncated to the spin $s\leq 2$ sector
for very special configurations. In other words, the solution
satisfies a highly non-local extension of scalar-coupled AdS
gravity (corresponding to an effective model given in
\cite{Varna}) as well as an infinite set of consistency conditions
for setting the higher-spin fields to zero. These conditions
reflect the invariance under an infinite-dimensional higher-spin
extension of $SO(3,1)$, that we denote by
\be hsl(2,C;\nu)\supset sl(2,C)\ ,\ee
with $\nu$-dependent generators given in \eq{hsl21} and
\eq{hsl22}. This motivates seeking solutions based on invariance
under other subgroups of $SO(3,2)$, which we shall analyze at the
linearized level in Section \ref{sec:5}.

The locally defined scalar and metric in \eq{intro:sol} are
components of a section of the higher-spin gauge-covariant master
fields, related to the section in the $\widetilde x^a$-chart by a
gauge transition defined in the overlap. The physical component
fields are thus related by more complicated ``duality''
transformations than standard reparameterizations. As a result, by
the $Z_2$-symmetry, they read the \emph{same} in $x^a$ and
$\widetilde x^a$-coordinates, and hence the scalar-field duality
transformation (in this particular background) assumes the form of
a fractional linear transformation, \emph{viz.}
\be \widetilde\phi(\widetilde x)\ =\ {\nu\phi(x)\over
\phi(x)-\nu}\ .\ee
The global solution therefore remains ``weakly coupled'' throughout
spacetime, and describes a Weyl-flat interpolation between two
asymptotically AdS$_4$ regions $\lambda^2x^2\sim 1$ and
$\lambda^2\widetilde x^2\sim 1$, via two $dS_3$-foliated domain
walls in $0<\lambda^2x^2,\lambda^2\widetilde x^2< 1$ and two FRW
spacetimes with $k=-1$ in the overlap region (one in the future and
one in the past).

As we shall discuss in Section \ref{sec:6}, the cosmological
interpretation of our solution requires the development of geometric
and algebraic tools suitable for the description of higher gauge
theory. Some of the properties of the solution aiming at such an
interpretation are reported in \cite{Varna}.


\scs{The Minimal Bosonic Model}\label{sec:2}


\scss{The Master-Field Equations}\label{sec:mfe}

To describe the higher-spin gauge theory based on the minimal
higher-spin algebra $hs(4)\supset SO(3,2)$, one introduces a set
of auxiliary coordinates $(z^\a,\bar{z}^{\ad})$ together with an
additional set of internal variables $(y_\a,\bar y_\a)$, that are
Grassmann-even $SL(2,C)$-spinor oscillators defined by the
associative product rules
\bea y_\a\star y_\b&=&y_\a y_\b+i\epsilon_{\a\b}\ ,\qquad
y_{\a}\star z_{\b}\ =\  y_{\a}z_{\b}-i\,\e_{\a\b}\ ,\label{osc1}\\[5pt] z_{\a}\star
y_{\b}&=& z_{\a}y_{\b}+i\,\e_{\a\b}\ , \qquad z_{\a}\star z_{\b}\
=\  z_{\a}z_{\b}-i\,\e_{\a\b} \ ,\label{osc2} \eea
where the juxtaposition denotes the symmetrized, \emph{i.e.}
Weyl-ordered, products. The hermitian conjugates
$(y_\a)^\dagger=\bar y_{\dot\a}$ and $(z_\a)^\dagger=\bar
z_{\dot\a}$ obey
\bea \bar y_{\dot\a}\star \bar y_{\dot\b}\ =\ \bar y_{\dot\a} \bar
y_{\dot\b}+i\epsilon_{\dot\a\dot\b}\ ,\qquad \bar z_{\dot\a}\star
\bar y_{\dot\b}\ =\ \bar z_{\dot \a} \bar y_{\dot\b}-
i\epsilon_{\dot\a\dot\b}\ ,\label{oscbar1}\\[5pt] \bar y_{\dot\a}\star \bar z_{\dot\b}\
=\ \bar y_{\dot\a} \bar z_{\dot\b}+i\epsilon_{\dot\a\dot\b}\
,\qquad \bar z_{\dot\a}\star \bar z_{\dot\b}\ =\ \bar z_{\dot\a}
\bar z_{\dot\b}-i\epsilon_{\dot\a\dot\b}\ .\label{oscbar2}\eea
Equivalently, in terms of Weyl-ordered functions
 \bea
 &&\widehat f(y,\bar y,z,\bar z)~\star~ \widehat g(y,\bar
y,z,\bar z)\label{star}\\[5pt]&=&\ \int \frac{d^2\xi d^2\eta d^2\bar\xi
d^2\bar\eta}{(2\pi)^4}~ e^{i\eta^\a\xi_\a+
i\bar\eta^{\dot\a}\bar\x_{\dot\a}} ~\widehat f(y+\xi,\bar y+\bar
\xi,z+\xi,\bar z-\bar \xi)~\widehat g(y+\eta,\bar y+\bar
\eta,z-\eta,\bar z+\bar \eta)\ ,\nn
 \eea
where the hats are used to indicate functions that depend on both
$(y,\yb)$ and $(z,\zb)$, while functions depending only on
$(y,\yb)$ shall be written without hats. The basic master fields
are differential forms in an extended spacetime with coordinates
$(x^M,z^\a,\zb^{\ad})$, namely a one-form
 \be
\widehat{A}\ =\ dx^M \widehat{A}_M(x,z,\bar z;y,\bar y) + dz^\a
\widehat{A}_\a(x,z,\bar z;y,\bar y)+
d\zb^{\ad}\widehat{A}_{\ad}(x,z,\bar z;y,\bar y)\ ,
 \ee
and a zero-form $\widehat{\Phi}=\widehat\Phi(x,z,\bar z;y,\bar
y)$.

The full higher-spin equations based on the above algebraic
structures were first given in \cite{vasiliev}. It can be shown
that parity invariance and manifest Lorentz invariance restricts
the possible interactions to two cases referred to as the minimal
Type A and Type B models, in which the scalar field is even and
odd under parity, respectively \cite{Vas:more,Sezgin:2003pt}. The
resulting master-field equations read
 \bea
\widehat{F}&=& \frac{i}4 b_1 dz^\a\wedge dz_\a~
\widehat{\Phi}\star \k + \frac{i}4 (b_1)^\star d\zb^{\ad}\wedge
d\zb_{\ad}~
\widehat{\Phi}\star \bar{\k}\ ,\la{c1}\\[5pt]
\widehat{D}\widehat{\Phi}&=& 0\ ,\la{c2}
 \eea
where $b_1=1$ and $b_1=i$ in the Type A and Type B models,
respectively, and the curvatures are defined as
\be \widehat{F}= d\widehat{A}+\widehat{A}\star\widehat{A}\
,\quad\quad \widehat{D}\widehat{\Phi}=d\widehat{\Phi}+[\widehat
A,\widehat \Phi]_\pi\ ,\ee
with
 \be
[\widehat f,\widehat g]_\pi\ =\ \widehat f\star\widehat
g-\widehat g\star\pi(\widehat f)\ ,\label{pitator}
 \ee
and the functions $\kappa$ and ${\bar\kappa}$ defined by
 \be
\k\ =\ \exp(iy^\a z_\a)\ ,\quad\quad \bar{\k}\ =\ \k^\dagger\ =\
\exp(-i\yb^{\ad}\zb_{\ad})\ ,
 \ee
have the salient properties
\be \kappa\star \widehat f(y,z)\ =\ \kappa \widehat f(z,y)\
,\qquad \widehat f(y,z)\star\kappa\ =\ \kappa \widehat f(-z,-y)\
.\label{kappa}\ee\be \kappa\star\widehat f\star\kappa\ =\
\pi(\widehat f)\ .\ee
In order to restrict to the minimal bosonic model one imposes
further kinematic conditions
\be {\tau}(\hA)= -\hA\ ,\quad \hA^\dagger =-\hA\ ,\quad\quad
{\tau}(\widehat\Phi)={\bar\pi}(\hF)\ ,\quad
\hF^{\dagger}=\pi(\hF)\ ,\la{hf2}\ee
where $\t$ is the anti-automorphism
\be \tau(\widehat f(y,\yb,z,\zb))\ =\ \widehat f(i y,i\yb,-i z,-i
\zb)\ ,\qquad \t(\widehat f\star\widehat g)\ =\ \t(\widehat
g)\star\t(\widehat f)\ ,\ee
and $\pi$ is the involutive automorphism
\be \pi(\widehat f(y,\yb,z,\zb))\ =\ \widehat f(- y,\yb,- z, \zb)\
=\ \kappa\star \widehat f\star\kappa \ ,\qquad\pi(\widehat
f\star\widehat g)=\pi(\widehat f)\star\pi(\widehat g)\
.\label{taupi}\ee
The gauge transformations are given by
\be \delta_{\widehat \e}\widehat A\ =\ \widehat D\widehat \e\
,\qquad \delta_{\widehat \e}\,\widehat\Phi\ =\ -[\widehat
\e,\widehat\Phi]_\pi\ .\label{gauge}\ee
The rigid, \emph{i.e.} $x$ and $Z$-independent, gauge parameters
form the Lie algebra

\be hs(4)\ =\ \left\{P(y,\bar y)\ :\quad \t(P)\ =\ P^\dagger\ =\
-P\ \right\}\ ,\ee
whose maximal finite-dimensional subalgebra is given by $SO(3,2)$,
with generators \eq{mab}.

To analyze the master-field equations \eq{c1} and \eq{c2}
formally, one starts from an ``initial'' condition
\be \Phi(x;y,\bar y)\ =\ \widehat \Phi|_{Z=0}\ ,\qquad
A_M(x;y,\bar y)\ =\ \widehat A_M|_{Z=0}\ ,\label{ic}\ee
and a suitable gauge condition on the internal connection, such as
\cite{Us:analysis}
\be \widehat A_{\a}|_{\Phi=0}\ =\ 0\ ,\label{physgauge1}\ee
and proceeds by obtaining the $Z$-dependence of the fields by
integrating the ``internal'' constraints, \emph{viz.}
\be \widehat D_\a\widehat\Phi\ =\ 0\ ,\qquad \widehat F_{\a\b}\ =\
-{ib_1\over 2}\e_{\a\b}\widehat\Phi\star\kappa\ ,\qquad \widehat
F_{\a\ad}\ =\ 0\ ,\qquad\widehat F_{\a M}\ =\ 0\ ,\ee
perturbatively in a $\Phi$-expansion, denoted by
\be \widehat \Phi\ =\ \Phi+\sum_{n=2}^\infty\widehat\Phi^{(n)}\
,\qquad \widehat A_\a\ =\ \sum_{n=1}^{\infty}\widehat A^{(n)}_\a\
,\qquad \widehat A_M\ =\ A_M+\sum_{n=1}^{\infty}\widehat
A_M^{(n)}\ .\label{pert}\ee
The $x$-space constraints $\widehat F_{MN}=0$ and $\widehat
D_M\widehat \Phi=0$ can then be shown to be perturbatively
equivalent to $\widehat F_{MN}|_{Z=0}=0$ and $D_M\widehat
\Phi|_{Z=0}=0$, that is
\be F_{MN}\ =\ -\sum_{p=1}^\infty\sum_{m+n=p}[\widehat
A_M^{(m)},\widehat A^{(n)}_{N}]_\star\ ,\qquad D_M\Phi =\
-\sum_{p=2}^\infty\sum_{m+n=p}[\widehat
A_M^{(m)},\widehat\Phi^{(n)}]_\pi\ ,\label{xsp}\ee
where $F_{MN}=2\partial_{[M}A_{N]}+[A_M,A_N]_\star$ and
$D_M\Phi=\partial_M\Phi+[A_M,\Phi]_\pi$. These equations
constitute a perturbatively Cartan-integrable system in $x$-space
provided that the \emph{full} $Z$-dependence is included at each
order in $\Phi$.

Since \eq{xsp} are written entirely in terms of differential forms
they are manifestly diffeomorphism invariant. In fact, they are
invariant under homotopy transformations, whereby coordinate
directions can be added and removed without affecting the physical
content. Thus, in case $x$-space is homotopic to a
four-dimensional space-time manifold with coordinates $x^\m$
($\m=0,1,2,3$), then one can without loss of generality formulate
\eq{xsp} directly on this four-manifold.


\scss{The Space-Time Field Equations}\label{sec:stfe}


In order to obtain the physical field equations on generally
covariant form, one first has to Lorentz covariantise \eq{xsp}. To
this end, one first identifies the full Lorentz generators acting
on the hatted master fields as follows
\be \widehat{M}_{\a\b}\ =\  \widehat  M^{(0)}_{\a\b} +\frac12 \,\{
\widehat S_\a,\widehat S_\b\}_*\ ,\label{Mhat}\ee
where $\widehat  M^{(0)}_{\a\b}=y_\a y_\b - z_\a z_\b$, and
\be \widehat{S}_\a\ =\ z_\a-2i\widehat{A}_\a\ .\label{salpha}\ee
One can show that \cite{Vas:star}
\bea \d_{L} \hF&\equiv &-[\widehat\epsilon_L,\widehat \Phi]_\pi\
=\ -[\widehat\e_0,\widehat{\Phi}]_{\star}\ ,\la{ll1}\w2 \d_{L}
\hA_\a &\equiv & \widehat D_\alpha\widehat\epsilon_L\ =\
-[\widehat\e_0, {\widehat A}_\a]_* +\L_\a{}^\b \hA_\b \
,\la{ll2}\w2 \d_{L} \widehat{A}_\m &\equiv&\widehat D_\mu\widehat
\epsilon_L\ =\
-[\widehat\e_0,\widehat{A}_\m]_{\star}+\left({1\over
4i}\partial_\m\L^{\a\b}\widehat M_{\a\b}-{\rm h.c.}\right)\
,\la{ll3}\eea
where $\widehat \epsilon_L=\frac1{4i}\Lambda^{\a\b}(x)\widehat
M_{\a\b}-({\rm h.c.})$ are the full parameters, and $\widehat
\epsilon_0=\frac1{4i}\Lambda^{\a\b}\widehat M^{(0)}_{\a\b}-({\rm
h.c.})$ are the parameters of canonical Lorentz transformations of
the $Y$ and $Z$ oscillators. The canonically transforming
component fields are thus obtained by $Y$ and $Z$-expansion of
$\widehat A_\a$, $\widehat \Phi$ and $\widehat A_\mu-({1\over 4i}
\omega_\mu{}^{\a\b}\widehat M_{\a\b}-{\rm h.c.})$ where
$\omega_\mu{}^{\a\b}$ is the Lorentz connection with $\d_L
\o_\m{}^{\a\b}=\partial_\m \L^{\a\b}+ \L^{\a\c}
\o_\m{\c}^{\b}+\L^{\b\c} \o_\m{\c}^{\a}$ (related conventions are
given in Appendix \ref{App:conv}). Hence, introducing
\bea \o_\m&=& {1\over 4i}\o_\m{}^{\a\b}M_{\a\b} -{\rm h.c.}\
,\qquad M_{\a\b}\ =\ y_\a y_\b\ ,\label{lorconn}\\
\widehat K_\mu&=&{1\over 4i}\omega_\mu{}^{\a\b}(\widehat
M_{\a\b}-M_{\a\b})-{\rm h.c.}\ ,\label{hatK}\eea
and using the gauge condition \eq{physgauge1} to simplify $K_\m
=\widehat K_\m|_{Z=0}$, the Lorentz covariant decomposition of the
master-gauge field $A_\mu$ reads
\bea A_\m&=&e_\m+\omega_\m+ W_\m +K_\m\ ,\label{Amu}\\ K_\m &=&
i\o_\m{}^{\a\b}
\left.(\widehat{A}_\a\star\widehat{A}_\b)\right|_{Z=0}-\mbox{h.c.}\label{k}\eea
where $e_\m$ is the vielbein\footnote{The vielbein here is given
in the higher-spin frame, where the torsion is non-vanishing. A
discussion of the Einstein frames is given in \cite{Varna}.}

\be e_\m\ =\ \frac1{2i}e_\m{}^{\a\ad}y_\a\yb_{\ad}\ ,\qquad
e_\m{}^{\a\ad}\ =\ -{\l\over 2}(\s_a)^{\a\ad} e_\m{}^a\ ,\la{e}
\ee
and $W_\m=W_\m(y,\yb)$ contains the higher-spin gauge fields.
Indeed, inserting \eq{Amu} into \eq{xsp}, the explicit Lorentz
connections cancel, and one obtains a manifestly Lorentz covariant
and space-time diffeomorphism invariant set of constraints
\cite{Vas:star,Us:analysis}.

To describe the higher-spin gauge symmetries in a generally
space-time covariant fashion, one proceeds using a
\emph{weak-field approximation} in which the higher-spin gauge
fields, the scalar field and the Weyl tensors, including that of
spin $2$, are treated as weak fields, while \emph{no approximation
is made for the vielbein and the Lorentz connection}. It can then
be shown that $\widehat D_\mu\widehat \Phi|_{Z=0}=0$ contains the
field equation for the physical scalar field
\be \phi\ =\ \Phi|_{Y=0}\ ,\label{scalar}\ee
and that $\widehat F_{\m\n}|_{Z=0}=0$ contains the field equation
for the metric
\be g_{\mu\nu}\ =\ e_\mu{}^{a}~e_{\nu a}\ =\
-2\l^{-2}e_\m{}^{\a\ad}e_{\n \a\ad}\ ,\ee
and a set of physical higher-spin fields given by the
doubly-traceless symmetric tensor fields of rank $s=4,6,...$,
given by
\be \phi_{\mu_1 \mu_2\dots \mu_{s}}\ =\ 2i~
e_{(\mu_1}^{\a_1\ad_1}\cdots
e_{\mu_{s-1}}^{\a_{s-1}\ad_{s-1}}\frac{\partial^{s-1}}{\partial
y^{\a_1}\cdots \partial
y^{\a_{s-1}}}\frac{\partial^{s-1}}{\partial \yb^{\ad_1}\cdots
\partial \yb^{\ad_{s-1}}} W_{\mu_s)}|_{Y=0}\ .\ee
The \emph{generally covariant physical field equations} are given
up to second order in weak fields by
\bea (\nabla^2+2\l^2)\,\phi &=& \left( \nabla^\mu P^{(2)}_\mu
-\frac{i\l}2 (\s^\mu)^{\a\ad} {\partial\over
\partial y^\a} {\partial\over\partial {\yb}^{\ad}}\,
P^{(2)}_\mu\right)_{Y=0}\ , \la{sfe3}\w4
\left(\s^{\m\n\r}\right)_\a{}^{\bd}\,{\cal R}_{\n\r\,\bd\ad}&=&
\left(\s^{\m\n\r}\right)_\a{}^{\bd}\, \left( {\partial
\over\partial {\yb}^{\bd}} {\partial \over \partial
{\yb}^{\ad}}J^{(2)}_{\n\r}\right)_{Y=0}\ , \la{efe3}\w4
\left(\s_{\m}{}^{\n\r}\right)^{\phantom{()}}_{(\a_1\phantom{\bd}\!\!\!}{}^{\!\!\bd}\,
{\cal F}_{\n\r\,\a_2\dots\a_{s-1})\bd  \ad_1 \dots\ad_{s-1}}&=&
\left(\s_{\m}{}^{\n\r}\right)^{\phantom{()}}_{(\a_1\phantom{\bd}\!\!\!}{}^{\!\!\bd}\left(
{\partial \over
\partial y^{\a_2}}\cdots {\partial\over \partial y^{\a_{s-1})}}
{\partial\over
\partial {\yb}^{\bd}}\cdots {\partial\over \partial {\yb}^{\ad_{s-1}
\phantom{\bd}\!\!\!}}\, J^{(2)}_{\n\r}\right)_{Y=0} \la{hsfe3}
\eea
where the source terms $P^{(2)}_\m$ and $J^{(2)}_{\m\n}$ are
provided in Appendix \ref{app:c}; ${\cal R}_{\m\n\ad\bd}$, which
is defined in \eq{rab}, is the self-dual part of the full
$SO(3,2)$-valued curvature $d\O+\O\star\O$ with $\O$ given in
\eq{Omega}; the higher-spin curvatures are defined by
($s=4,6,8,..$)
\bea {\cal F}_{\n\r\,\a_2\dots\a_{s-1}\bd \cd \ad_2
\dots\ad_{s-1}} &=& 2\nabla_{[\n }W_{\r
]\a_2\dots\a_{s-1}\bd\cd\ad_2\dots\ad_{s-1}} \nn\w2
&&-(s-2)(\s_{\n\r}\s_\m)_{(\a_2}{}^{\dot d}\,
W_{\m\,\a_3\dots\a_{s-1})\bd\cd{\dot\d}\ad_2\dots\ad_{s-1}} \nn\w2
&& -s (\s_\m\s_{\n\r})_{(\bd}{}^\c\,
W_{\m\,\c\a_2\a_3\dots\a_{s-1}\cd\ad_2\dots\ad_{s-1})}\ ; \la{f1}
\eea
and the covariant derivatives in \eq{sfe3} and \eq{f1} are given
by $\nabla=d+\o$ with $\o$ being the canonical Lorentz connection.
The expression \eq{f1} contains the auxiliary gauge fields
$W_{\m\a(s-2)\ad(s)}$ and $W_{\m\a(s-3)\ad(s+1)}$, of which the
latter drops out from \eq{hsfe3}, while the former can be
expressed explicitly in terms of the physical fields using
curvature-dressed covariant derivatives $\widehat\nabla$, as
explained in \cite{Us:analysis}.


\scs{Construction of Solutions}\label{sec:3}



\scss{Perturbative Space-Time Approach}\label{sec:wfe}


In the first order of the weak-field expansion, it is consistent
to truncate the higher-spin field equations to that of pure
Einstein gravity with cosmological constant plus a free scalar
field in a fixed background metric, \emph{viz.}
\be \bar R_{\mu\nu}+3\lambda^2~\bar g_{\mu\nu}\ =\ 0\ ,\qquad
(\bar \nabla^2+2\l^2)\bar\phi\ =\ 0\ .\label{1st}\ee
These equations are self-consistent, though they do not derive
from an action. They provide a good approximation if $\bar\phi$,
the spin-$2$ Weyl tensor $\bar \Phi_{\m\n\r\s}$ and all their
derivatives are small. In general, such solutions may describe
spacetimes that are not asymptotically conformally flat in any
region.

The higher-order corrections from the weak-field expansion yields
a perturbative expansion in $\bar\phi$ and $\bar \Phi_{\m\n\r\s}$
of the form
\be g_{\mu\nu}\ =\ \bar g_{\mu\nu}+\sum_{n=2}^\infty
g^{(n)}_{\m\n}\ ,\qquad \phi\ =\ \bar \phi+\sum_{n=2}^\infty
\phi^{(2)}\ ,\qquad W_\m\ =\ \sum_{n=2}^\infty W^{(n)}_\m\ .\ee
The second-order corrections can be determined from \eq{sfe3},
\eq{efe3} and \eq{hsfe3}, where the higher-spin gauge fields and
Weyl tensors do not contribute to $P^{(2)}_\m$ and
$J^{(2)}_{\m\n}$ to that order. In this sense, exact solutions to
ordinary Einstein gravity with cosmological constant may be
embedded into higher-spin gauge theory, albeit that finding the
exact solutions in closed form amenable to studies of salient
properties is highly non-trivial. Moreover, there are subtleties
related to boundary conditions as well as the convergence of the
perturbative expansion.

The higher-spin field equation \eq{hsfe3} reads ($s=4,6,\dots$)
\be \bar{\cal F}_{\m_1\dots \m_s}\ =\ {\cal O}((\mbox{weak
fields})^2)\ ,\label{frons}\ee
where the Fronsdal-like operator $\bar{\cal F}_{\m_1\dots \m_s}$
is covariantised using the background metric $\bar g_{\m\n}$. The
first-order truncation of \eq{frons} is not
\emph{self-consistent}, \emph{i.e.} incompatible with linearized
higher-spin gauge symmetry, unless $\bar \Phi_{\m\n\r\s}=0$,
\emph{i.e.} $\bar g_{\m\n}=g_{(0)\m\n}$ where the subscript $(0)$
denotes the AdS$_4$ background with curvature given by \eq{ads}.
Thus, the proper way to include higher-spin fields $\bar
\phi_{\m_1\dots \m_s}$ into the first order is to solve
\be \bar{\cal F}_{\m_1\dots\m_s}\ =\ 0\ ,\ee
including all gauge artifacts, and impose the gauge conditions
order-by-order in weak-field expansion using the fully consistent
higher-spin field equation \eq{hsfe3}.


In order to switch on higher-spin fields, it is therefore
convenient to consider solutions in which the full Weyl zero-form
$\Phi$ asymptotes to zero in some region of spacetime. We shall
refer to such solutions as \emph{asymptotically Weyl-flat
solutions}.


The perturbative approach to these solutions is self-consistent in
the sense that the linearized twisted-adjoint zero-form
$C=C(x;y,\yb)$ obeying $dC+[\O_{(0)},C]_\pi=0$, where $\O_{(0)}$
is the AdS$_4$ connection, vanishes on the boundary. To
demonstrate this, one writes the linearized equation as
 \be
 \nabla_{(0)\mu} C+{\l\over 2i}e_\m{}^a\{P_a,C\}_\star\ =\ 0\
 ,\label{linc}
 \ee
and expands $C$ and \eq{linc} in $y$ and $\yb$, which shows that
$C$ contains linearized Weyl tensors $C_{a(s),b(s)}$ obeying
($s=2,4,\dots$)
\be e^{a}_{(0)}\wedge e^b_{(0)} \wedge e^c_{(0)}
~\nabla_{(0)a}C_{b\mu(s-1),c\nu(s-1)}\ =\ 0\ ,\qquad
\nabla^\r_{(0)} C_{\r\mu(s-1),\nu(s)}\ =\ 0\ .\label{wfe}\ee
These equations are in fact valid in AdS$_D$. The resulting
mass-shell condition reads
\be \left[\nabla^2_{(0)}+2(s+D-3)\l^2\right]C_{\mu(s),\nu(s)}\ =\
0\ .\ee
Setting $s=0$ one obtains formally the correct scalar-field
equation,
\be (\nabla^2_{(0)}+2(D-3)\l^2)\varphi\ =\ 0\ ,\label{freesf}\ee
where $\varphi=C|_{Y=0}$ in $D=4$. Splitting $x^\mu\rightarrow
(x^i,r)$, and using Poincar\'e coordinates,
 \be
 ds^2_{(0)}\ =\ {1\over\lambda^2r^2} \left(dr^2+dx^2\right)\ ,\quad
 \C^r_{(0)ij}= {1\over r}\eta_{ij}\ ,\quad \C^j_{(0)ir} = -{1\over r} \d_i^j\ ,
 \quad \C^r_{(0)rr}= -{1\over r}\ ,\label{Poincare}
 \ee
one finds that the component fields $C_{i(s),j(t)r(s-t)}$ with
$s\geq t\geq 1$, and where the indices are curved, are given by
curls of $C_{i(s),r(s)}$, taken using $r\partial_j$, that in their
turn obey
 \be
 \left[\left(\partial_r+{2-D\over r}\right)
 \left(\partial_r+{2s\over r}\right)+{s^2+(D-1)s+2(D-3)\over
 r^2}+\eta^{ij}\partial_i\partial_j\right]C_{i(s),r(s)}\ =\ 0\ .
 \ee
Thus, the linearized spin-$s$ Weyl tensor $C_{a(s),b(s)}$ consists
of two sectors $C^{(\pm)}_{a(s),b(s)}$ with scaling behavior given
by ($s=0,2,4,\dots$)
 \be
 C^{(\pm)}_{a(s),b(s)}\ \sim\ r^{\beta^{\pm}_s}\ ,\qquad \b^{\pm}_s\ =\ s+{D-1\over 2}\pm
 {D-5\over 2}\ .
 \label{rpm}
 \ee

At higher orders of the weak-field expansion, and due to the
higher-derivative interactions hidden in the $\star$-products in
$\widehat D_\m\widehat\Phi|_{Z=0}=0$, the corrections to the
spin-$s$ Weyl tensor $\Phi$ may in principle contain lower-spin
constructs with a total scaling weight less than $\b^-_s$. We
shall not analyze the nature of these corrections in any further
detail here, but hope to return to this interesting issue in a
future work.

Another non-local effect induced via the $Z$-space dependence, is
that the Lorentz covariantisations in $K_\m$, defined in \eq{k},
may remain finite in the asymptotic region, despite the naive
expectation that the scaling of $(\widehat A_{(\a}\star\widehat
A_{\b)})|_{Z=0}$, which is of order $\Phi^2$, should over-power
that of $\o_\m{}^{\a\b}$. While this is a challenging problem to
address in its generality, we shall find that already the
relatively simple case of the $SO(3,1)$-invariant asymptotically
Weyl-flat solution exhibits an interesting phenomenon whereby
$K_\m$ generates a finite Weyl rescaling and contorsion in the
asymptotic region.

Clearly, the weak-field expansion, which is naturally geared towards
dressing up solutions to \eq{1st}, is going to be far from efficient
in dealing with general asymptotically Weyl-flat solutions.
Especially when the starting point is not a solution to \eq{1st}, it
is appropriate to develop an alternative approach to solving the
basic master-field equations \eq{c1} and \eq{c2} in which one makes
a maximum use of the fact that the \emph{local} $x$-dependence is a
gauge choice. As we shall see next, the Z-space approach indeed does
exploit this fact, and provides a powerful framework for finding
exact solutions to higher-spin field equations.


\scss{The $Z$-Space Approach}\label{sec:zspace}


In order to construct solutions one may consider the $Z$-space
approach \cite{Bolotin:1999fa} in which the constraints in
spacetime, \emph{viz.}
 \be
 \widehat F_{\m\n}\ =\ 0\ ,\qquad \widehat F_{\m\a}\ =\ 0\
,\qquad \widehat D_\m \widehat\Phi\ =\ 0\ ,\label{xsp2}
 \ee
are integrated in simply connected space-time regions given the
space-time zero-forms at a point $p$,
\be
 \widehat \Phi'\ =\ \widehat \Phi|_{p}\ ,\qquad \widehat
A'_\a\ =\  \widehat A_\a|_{p}\ ,\label{phiprime}
 \ee
and expressed explicitly as
 \be
\widehat A_\mu\ =\ \widehat L^{-1}\star \partial_\mu \widehat L\
,\qquad \widehat A_\a\ =\ \widehat L^{-1}\star (\widehat A'_\a+
\partial_\a) \widehat L\ ,\qquad \widehat \Phi\ =\ {\widehat L}^{-1}\star
\widehat\Phi'\star \pi(\widehat L)\ ,\label{Leq}
 \ee
where $\widehat L=\widehat L(x,z,\bar z;y,\yb)$ is a gauge
function, and
\be \widehat L|_{p}\ =\ 1\ ,\qquad \partial_\m{\widehat A}'_\a\ =\
0\ ,\qquad
\partial_\m\widehat\Phi'\ =\ 0\ .\ee
The remaining constraints in $Z$-space, \emph{viz.}
\bea \widehat F'_{\a\b}&\equiv&2\partial_{[\a}\widehat
A'_{\b]}+[\widehat A'_\a,\widehat A'_\b]_\star\ =\ \ -\frac{ib_1}2
\e_{\a\b}\widehat\Phi'\star \kappa\ ,\label{z1}\\[5pt]\widehat F'_{\a\bd}&\equiv&
\partial_\a\widehat A'_{\bd}-\partial_{\bd}\widehat
A'_{\a}+[\widehat A'_\a,\widehat A'_{\bd}]_{\star}\ =\ 0\ ,\label{z2}\\[5pt]
\widehat D'_\a\widehat \Phi'&\equiv& \partial_\a\widehat
\Phi'+\widehat A'_\a\star\widehat\Phi'+\widehat
\Phi'\star\pi(\widehat A'_\a)\ =\ 0\ ,\label{z3}\eea
must then be solved given an initial condition
 \be
C'(y,\bar y)\ =\ \widehat\Phi'|_{Z=0}\ ,\label{C}
 \ee
and fixing a suitable gauge for the internal connection. The
natural choice is
 \be
 \widehat A'_\a|_{C'=0}\ =\ 0\ ,\label{physgauge}
 \ee
whose compatibility with \eq{physgauge1} requires
\be \widehat L|_{C'=0}\ =\ L(x;y,\yb)\ ,\ee
that is, the gauge function cannot depend explicitly on the
$Z$-space coordinates. In what follows, we shall assume that
\be \widehat L\ =\ L(x;y,\yb)\ .\ee
The gauge fields can then be obtained from \eq{Amu}, \emph{viz.}
\be e_\m+\o_\m+W_\m\ =\ L^{-1}\partial_\m L-K_\m\
,\label{gf}\ee
where
\be K_\m\ =\ \widehat K_\m|_{Z=0}\ ,\qquad \widehat K_\mu\ =\
i\o_\m{}^{\a\b}L^{-1}\star \widehat A'_\a\star \widehat A'_\b\star
L\ .\ee
Hence, the gauge fields, including the metric, can be obtained
algebraically without having to solve any differential equations
in spacetime.

The local representatives in two overlapping simply connected
regions with gauge functions $L$ and $\tilde L$, are related on
the overlap via a gauge transformation with transition function
\be g\ =\ L^{-1}\star \tilde L\ .\ee
The resulting transformations of the physical fields, \emph{viz.}
\bea \widetilde e_\m+\widetilde \o_\m+\widetilde W_\m&=&
g^{-1}\star\left[e_\mu+\omega_\mu+W_\mu+K_\m-g\star(g^{-1}\star
\widehat K_\mu\star g)|_{Z=0}\star g^{-1}+\partial_\m\right]\star
g\ ,\quad\label{trans1}\\
\widetilde\phi&=&(g^{-1}\star \widehat \Phi\star g)|_{Z=0}\
.\label{trans2}\eea
which are a manifest symmetry of the full constraints in $x$ and
$Z$-space, are a non-trivial symmetry from the point of view of
the generally covariant space-time field equations contained in
$\widehat F_{\m\n}|_{Z=0}=0$ and $\widehat D_\mu\widehat
\Phi|_{Z=0}=0$ (whose general covariance corresponds to
field-dependent gauge parameters).

To describe asymptotically Weyl-flat solutions, one chooses the
gauge function to be a parameterization of the coset
\be L(x;y,\yb)\in {SO(3,2)\over SO(3,1)}\ .\label{l} \ee
By construction, its Maurer-Cartan form
 \be
 L^{-1}\star dL\ =\ \O_{(0)}\ =\ e_{(0)}+\omega_{(0)}\ ,
 \label{ads4}
 \ee
which means that the gauge fields defined by \eq{gf} are given
asymptotically by the AdS$_4$ vacuum plus corrections from $K_\m$.
The latter are higher order in the $C'$-expansion\footnote{Using
the gauge function \eq{wL}, the spin-$s$ sector in
$C(x;y,\yb)=L^{-1}(x)\star C'(y,\yb)\star L(x)$ scales like
$h^{2(1+s)}$, \emph{i.e.} $C'$ contains the regular boundary data
scaling like $r^{\b^-_s}$ in the notation of \eq{rpm} with $r\sim
h^2$.}, but may nonetheless contribute in the asymptotic region to
the leading dependence on the radial coordinate defined in
\eq{Poincare}. A concrete exemplification of this subtlety is
provided by the asymptotic behavior of the scale factors in the
$SO(3,1)$-invariant solution discussed at the end of Section
\ref{sec:spacetime}.


\scss{On Regular, Singular and Pseudo-Singular Initial
Conditions}\label{sec:sing}


The perturbative approach to \eq{z1}--\eq{z3} yields a solution of
the form
\be \widehat \Phi'\ =\ C'+\sum_{n=2}^\infty
\widehat\Phi^{\prime(n)}\ ,\qquad \widehat A'_\a\ =\
\sum_{n=1}^\infty \widehat A^{\prime(n)}_\a\ ,\label{pe}\ee
where the superscript indicate the order in $C'$. We shall refer
to $C'$ as a \emph{regular initial condition} provided that its
$\star$-product self-compositions, \emph{viz.}

\bea C'_{(2n)}&=& (C'\star \pi(C'))^{\star n}\ ,\label{c2n}\\
C'_{(2n+1)}&=& (C'\star \pi(C'))^{\star n}\star C'\
,\label{c2n1}\eea
are regular functions. The second-order corrections \eq{a2} and
\eq{Phi2} contain the $\star$-product composition
\be \left(C'(-t z,\yb)e^{ityz}\right)\star C'(y,\yb)\ =\
\kappa\star \left(C'(-t y,\yb)e^{i(1-t)yz}\right)\star C'(y,\yb)\
,\label{stru}\ee
where $t$ is the auxiliary integration parameter used to present
the first-order correction \eq{a1} to the internal connection. If
$C'$ is regular, then it follows that \eq{stru} is a regular
function of $(Y,Z)$ with $t$-dependent coefficients that are
analytic at $t=1$. We shall assume analyticity also at $t=0$,
where \eq{stru} involves only anti-holomorphic contractions
resulting in a ``softer'' composition that should not blow up.
Under these assumptions there exists an open contour from $t=0$ to
$t=1$ along which \eq{stru} is analytic, that can then be used in
\eq{a1} to produce a well-defined second order correction. This
argument extends to higher orders of perturbation theory, and
hence regular initial data yields perturbative corrections that
can be presented using open integration contours
\cite{Us:analysis}.

If $C'$ is not regular, and \eq{stru} has an \emph{isolated}
singularity at $t=1$, we shall refer to $C'$ as a \emph{singular
initial condition}. In this case, a well-defined perturbative
expansion can be obtained by circumventing the singularity by
using a closed contour $\gamma$ as follows,
\be d'\oint_\c {dt\over 2\pi i}\log\left({t\over 1-t}\right)
Z^{\underline\b} f_{\underline\b}(t Z)\ =\
dZ^{\underline\a}f_{\underline\a}(Z)\ ,\label{oint1}\ee
and
\be d'\oint_\c {dt\over 2\pi i}\left(t\log{t\over
1-t}-1\right)Z^{\underline\a}dZ^{\underline\b}f_{\underline{\a\b}}(t
Z)\ =\ \frac12 dZ^{\underline\a}\wedge
dZ^{\underline\b}f_{\underline{\a\b}}(Z)\ ,\label{oint2}\ee
where $Z^{\underline\a}=(z^\a,\zb^{\ad})$,
$d'=dZ^{\underline\a}\partial_{\underline\a}$, $d'f_1=d'f_2=0$ and
$\c$ encircles the branch cut from $t=0$ to $t=1$. The resulting
closed-contour presentation of the perturbative solution can of
course also be used in the case of regular initial data, in which
case $\gamma$ can be collapsed onto the branch-cut as to reproduce
the open-contour presentation.

Singular initial conditions may arise from imposing symmetry
conditions on $C'$, as will be exemplified in Section
\ref{sec:symm} in cases where the symmetry refers to unbroken
space-time isometries. Here the initial conditions are
parameterized elementary functions of the oscillators, \emph{e.g.}
combinations of exponentials of the bilinear translation generator
$P_a$ contracted with fixed vectors, that become singular for
particular choices of the parameters.

Another interesting type of irregular initial data arises in the
five-dimensional and seven- dimensional higher-spin models based
on spinor oscillators \cite{5d,5dn4,7d} and the $D$-dimensional
model based on vector oscillators. Here the initial conditions are
regular functions multiplied with \emph{singular projectors} whose
role is to gauge internal symmetries in oscillator space in order
that the master-field equations contain dynamics. These symmetries
are generated by bilinear oscillator constructs, $K$, and the
singular projectors are special functions with auxiliary integral
representations, given schematically by $\int_0^1 ds f(s) e^{sK}$,
such that the analogs of the self-compositions \eq{c2n} and
\eq{c2n1} now contain \emph{logarithmic divergencies arising in
corners of the auxiliary integration domain}. Due to the
projection property, these divergencies can be factored out and
written as \cite{Sagnotti:2005ns}
\be \int_0^1 {ds\over 1-s}g(s)\ ,\ee
where $g(s)$ is analytic and non-vanishing at $s=1$. One can now
argue that the perturbative expansion gives rise to analogs of
\eq{stru} resulting in regular functions of $(Y,Z)$ with
$t$-dependent coefficients involving pre-factors of the form
\be \int_0^1 {ds\over 1-t s}g(s)\ ,\ee
that are logarithmically divergent at $t=1$. Thus, the
open-contour presentation results in well-defined second-order
perturbations containing pre-factors of the form
\be \int_0^1\int_0^1{dsdt\over 1-st}g(s)\ ,\ee
while the closed-contour presentation, which requires analyticity
on closed curve encircling $t=1$, does not apply since the
singularity at $t=1$ is not isolated. We shall therefore refer to
initial conditions of this type as \emph{pseudo-singular initial
conditions}.

The above analysis indicate that the (pseudo-)singular nature of
initial conditions is an artifact of the naive application of the
$\star$-product algebra to \eq{c2n1}, while the actual
perturbative expansion in $Z$-space involves a point-splitting
mechanism that softens the divergencies. We plan to return with a
more conclusive report on these important issues in a forthcoming
paper.


\scss{Zero-Form Curvature Invariants}\label{sec:inv}


In unfolded dynamics, the local degrees of freedom are dual to the
twisted-adjoint element $C'(y,\yb)$ defined by \eq{C}, consisting
of all gauge-covariant derivatives of the physical fields
evaluated at a point in spacetime. At the linearized level, $C'$
is gauge covariant, which means that if $C'=g^{-1}\star \tilde
C'\star \pi(g)$, with $g$ a group element generated by $hs(4)$,
then $C'$ and $\tilde C'$ give rise to gauge-equivalent solutions.

To distinguish between gauge-inequivalent solutions at the full
level, we propose the following invariants
 \be
 {\cal C}^{\pm}_{2p}\ =\ {\cal N}_\pm \widehat Tr_\pm
 \widehat {\cal C}_{2p} \ ,\label{Cninv}\ee
for $p=1,2,\dots$, where
\be \widehat {\cal C}_{2p}\ =\
 \underbrace{\widehat \Phi\star
\kappa\star\cdots\star\widehat \Phi\star \kappa}_{\mbox{$2p$
times}}\ =\ (\widehat \Phi\star \pi(\widehat\Phi))^{\star p}\
;\label{Cn}
 \ee
the full traces are defined by
\be \widehat Tr_+\widehat f\ =\ \int {d^2y d^2\yb d^2z d^2\zb\over
(2\pi)^4} \widehat f(y,\yb,z,\zb)\ ,\qquad \widehat Tr_-\widehat
f\ =\ \widehat Tr_+(\widehat f\star \kappa\bar\kappa)\ ;\ee
and the normalizations ${\cal N}_\pm$ are given by
\be {\cal N}_-\ =\ 1\ ,\qquad {\cal N}_+\ =\ {1\over {\cal V}}\
,\ee
where ${\cal V}$ is the volume of $Z$-space. As separate
components of $\Phi$ vary over spacetime, the net effect is that
the invariants ${\cal C}^\pm_{2p}$ remain constant, though they
may diverge for specific solutions.

Let us motivate the above definitions. To begin with, it follows
from \eq{star} and \eq{kappa} that the full traces obey
\be \widehat Tr_\pm(\widehat f(Y,Z)\star\widehat g(Y,Z))\ =\
\widehat Tr_\pm(\widehat g(\pm Y,\pm Z)\star \widehat f(Y,Z))\
,\label{Trcyclic}\ee
and
\be \widehat Tr_-f(Y)\ =\  Tr_- f(Y)\ ,\label{f0}\ee
where the reduced traces $Tr_\pm f(Y)$ are defined by\footnote{The
odd trace $Tr_-$ for the Weyl algebra was originally introduced in
\cite{vas:supertrace}.}
\be Tr_+ f(Y)\ =\ \int {d^4Y\over (2\pi)^2}  f(Y)\ ,\qquad Tr_-
f(Y)\ =\ f(0)\ .\ee
Moreover, from $\widehat D_\m\widehat \Phi=0$, which is equivalent
to $\partial_\m (\widehat \Phi\star\k)=-[\widehat
A_\m,\widehat\Phi\star\kappa]$, it follows that
\be \partial_\m\widehat{\cal C}_q \ =\ -[\widehat
A_\m,\widehat{\cal C}_q]_\star\ ,\qquad \widehat{\cal C}_q\ =\
(\widehat\Phi\star\kappa)^{\star q}\ ,\ee
for any positive integer $q$, which together with \eq{Trcyclic}
and the fact that $\widehat A_\m$ is an even function of all the
oscillators implies that formally ($q=1,2,\dots$)
\be d \,\widehat Tr_\pm\widehat{\cal C}_q\ =\ 0\ .\ee
This property can be made manifest by going to primed basis using
\eq{Leq}.

Expanding perturbatively,
\be \widehat {\cal C}_{q}\ =\ (\Phi\star\kappa)^{\star
q}+\sum_{n=q+1}^\infty \widehat{\cal C}_{q}^{(n)}\ ,\ee
and taking $q=2p$, one finds that the leading contribution to the
charges are given by
\be {\cal C}_{2p}^{-(2p)}\ =\ Tr_-(\Phi\star\pi(\Phi))^{\star p}\
,\ee
where we have used \eq{f0}, and
\be  {\cal C}^{+(2p)}_{2p}\ =\ Tr_+ (\Phi\star\pi(\Phi))^{\star
p}\ ,\ee
where we have formally factored out and cancelled the volume of
$Z$-space. For $q=2p+1$ one finds that the leading order
contribution to $\widehat Tr_\pm\widehat{\cal C}_{2p+1}$ diverges
like $\int d^2z$ or $\int d^2\zb$, whose regularization we shall
not consider here. The higher-order corrections to the charges,
may require additional prescriptions, and, as already mentioned,
they need not be well-defined in general.

The form of the leading contribution ${\cal C}_{2p}^{-(2p)}$
suggests that the full charges ${\cal C}_{2p}^{-}$ are
well-defined for general regular initial data. Moreover, these
charges can be rewritten on a more suggestive form using \eq{c1},
where $dz^\a\wedge dz_\a=-2d^2z$, and
$\pi(\widehat\Phi)=\bar\pi(\widehat\Phi)$, which imply
\be \widehat F\star \widehat F\star \widehat{\cal C}_q\ =\
-{1\over 2} d^2z \wedge d^2\bar z\,\widehat{\cal
C}_{q+2}\star\kappa\bar\kappa\ ,\ee
so that
\be {\cal C}^-_{2p}\ =\ -Tr_+ \left[{1\over 2\pi^2}\int \widehat
F\star \widehat F\star (\widehat
\Phi\star\pi(\widehat\Phi))^{\star (p-1)}\right]\ ,\label{hint}\ee
which can be used to show that the charges can be written as total
derivatives in $Z$ order by order in the $\Phi$-expansion.


\scs{An $SO(3,1)$-Invariant Solution}\label{sec:4}


\scss{The Ansatz}

To find $SO(3,1)$-invariant solutions we use the $Z$-space approach
based on \eq{Leq}. It is convenient to use a Lorentz-covariant
parameterization of the gauge function, \emph{viz.}
\cite{Bolotin:1999fa} $$L(x;y,\yb)=\exp [i f(x^2) x^{\a\ad}
y_\a\yb_{\ad}+r(x^2)]$$ with $x^{\a\dot \a}=(\s_a)^{\a\dot\a}x^a$
and $x^2=x^a x_a$. The function $r(x^2)$ is fixed by demanding
$\t(L)=L^{-1}$, so that $L\in SO(3,2)$ and hence $L^{-1}\star dL$
describes the $AdS_4$ vacuum solution\footnote{To exhibit $L\in
SO(3,2)$ one can write $L(x;y,\yb)=\exp_\star (i{{\rm artanh}
\sqrt{1-h\over 1+h}\over \sqrt{1-h^2}}\lambda
x^{\a\ad}y_\a\yb_{\ad})$ where $\exp_\star A=1+A+\frac 12 A\star
A+\cdots$. }. Using $L^{-1}=(1-f^2 x^2)^2 \exp[ -i f(x^2) x^{\a\ad}
y_\a\yb_{\ad}-r(x^2)]$ one finds $r=\log (1-f^2 x^2)$. A convenient
choice of $f$ is \cite{Bolotin:1999fa}
\bea L(x;y,\yb)&=&{2h\over 1+h} \exp\left[{i\lambda
x^{\a\dot\a}y_\a \bar y_{\dot\a}\over 1+h}\right] \ ,\qquad
\lambda^2 x^2\ <\ 1\ , \label{wL} \eea
where
\be x_{\a\ad}\ =\ (\s^a)_{\a\ad} x_a\ ,\qquad h\ =\
\sqrt{1-\lambda^2x^2}\ ,\qquad x^2\ =\ x^a x_a\ee
corresponding to the vierbein and Lorentz connection
\be e_{(0)}{}^{\a\ad}\ =\ -{\l(\s^a)^{\a\ad}dx_a\over h^2}\
,\qquad \o_{(0)}{}^{\a\b}\ =\ - {\l^2(\s^{ab})^{\a\b} dx_a
x_b\over h^2}\ ,\label{adseo}\ee
that in turn gives
\be ds^2_{(0)}\ =\ {4 dx^2\over (1-\lambda^2x^2)^2}\ ,\ee
which one identifies as the metric of AdS$_4$ spacetime with inverse
radius $\l$ given in stereographic coordinates. The inversion
$x^\m\rightarrow -x^\mu/(\l^2x^2)$ maps the space-like regions $0<
\l^2x^2<1$ and $\l^2x^2>1$ into each other and the boundary
$\l^2x^2=1$ onto itself. The future and past time-like regions are
mapped onto themselves, with distant past and future, where
$\l^2x^2\rightarrow-\infty$, sent to the past and future light
cones, respectively. The two space-like and the two time-like
regions provide a global cover of $AdS_4$ spacetime, so that beyond
the distant past and future lies the space-like region $\lambda^2
x^2>1$, as can be seen more explicitly using the global
parametrization $ds^2=-(1+r^2)dt^2+(1+r^2)^{-1}dr^2+r^2d\Omega_2^2$
where $x^2=2(1+\sin t\sqrt{1+r^2})^{-1}$. Thus, a global description
can be obtained using the additional gauge function,
\be \widetilde L \ =\ L(\widetilde x;y,\bar y)\ ,\qquad
\lambda^2\widetilde x^2\ <\ 1\ ,\label{wtL}\ee
and declaring the overlap with $L(x;y,\yb)$ to be given by
\be \widetilde x^a\ =\ -{x^a\over \lambda^2 x^2}\ ,\qquad
\lambda^2 x^2\ =\ 1/(\lambda^2\widetilde x^2)\ <\ 0\ ,\ee
leading to the transition function $\widetilde L^{-1}\star L$. The
$Z_2$-symmetry implies that the local representatives of the full
solution will be given by the \emph{same} functions, with
$x^a\leftrightarrow \widetilde x^a$, as we shall discuss in more
detail in Section \ref{sec:spacetime}\footnote{One can describe
anti-de Sitter spacetime using the globally well-defined gauge
function $L=(1-\l^2x^2) \exp i\l x^{\a\ad} y_\a\yb_{\ad}$ where
$\l^2 x^2\neq 1$. The corresponding $SO(3,1)$-invariant solution has
Lorentz connection and vierbein given by \eq{sol:lorconn} and
\eq{sol:vierbein} with $a^2$ replaced by $x^2$ in \eq{Q}. These
expressions are ill-defined for $|x^2|>1$, which means that a
globally well-defined solution still requires another coordinate
patch. }.

A particular type of $SO(3,1)$-invariant solutions can be obtained
by imposing
\bea [\widehat M'_{\a\b},\widehat \Phi']_\pi&=&0\ ,\label{inv1}\\
\widehat D'_\a\widehat M'_{\b\c}&=&0\ ,\label{inv2}\eea
where $\widehat M'_{\a\b}$ are the full Lorentz generators defined
in \eq{Mhat} and given in the primed basis, and obeying
\eq{ll1}--\eq{ll3}. Eq. \eq{inv1} combined with \eq{ll1} imply
that
\be \widehat \Phi'\ =\ \widehat \Phi'(u,\bar u)\ ,\ee
where
\be u\ =\ y^\a z_\a\ ,\qquad \bar u\ =\ u^\dagger\ =\ \bar y^{\dot
\a} \bar z_{\dot \a}\ ,\ee
and $(\widehat\Phi'(u,\bar u))^\dagger=\widehat\Phi'(u,\bar u)$.
Moreover, from \eq{inv2} combined with \eq{ll2} and the
$\tau$-invariance condition on $\widehat A_\alpha$, it follows
that
\be \widehat A'_\alpha\ =\ z_\a~A(u,\bar u)\ ,\qquad \widehat
S'_\alpha\ =\ z_\a~S(u,\bar u)\ ,\qquad S=1-2i A\
.\label{ansatz}\ee

We next turn to the exact solution of the $Z$-space equations
\eq{z1}--\eq{z3}.


\scss{Solution of $Z$-Space Equations}


The internal constraints $\widehat F'_{\a\dot\a}=0$ and $\widehat
D'_\alpha \widehat \Phi'=0$ are solved by
\be S(u,\bar u)\ =\ S(u)\ ,\qquad \widehat\Phi'(u,\bar u)\ =\
{\nu\over b_1}\ ,\label{sol1}\ee
where $\nu/b_1$ is a real constant, so that $\nu$ is real in the
Type A model and purely imaginary in the Type B model.

The remaining constraint on $\widehat F'_{\a\b}$, given by
\eq{z1}, now takes the form
\be [\widehat S^{\prime\alpha},\widehat S'_\a]_\star\ =\ 4i(1- \nu
e^{iu})\ .\label{SSnu}\ee
To solve this constraint, following \cite{Prokushkin:1998bq}, we
use the integral representation
\be S(u)\ =\ \int_{-1}^1 ds~ m(s) ~e^{\frac{i}2(1+s)u}\ ,\ee
where the choice of contour is motivated by the relation
\bea&& (z_\a e^{\frac{i}2(1+s)u})\star (z_\b
e^{\frac{i}2(1+s')u})\
\nn\\[5pt]&=&
\left(-i\e_{\a\b}-\frac14[y-z+s'(y+z)]_\a[y+z+s(y-z)]_\b\right)
e^{ \frac{i}2(1-s s')u}\ ,\label{zz}\eea
which induces the map $(s,s')\mapsto -ss'$ from $[-1,1]\times
[-1,1]$ to $[-1,1]$. As a result \eq{SSnu} becomes
\be \int_{-1}^1 ds \int_{-1}^1 ds' \left[1+\frac{i}4(1-s
s')u\right]~m(s)~m(s')~e^{\frac{i}2(1-s s')u}\ =\ 1-\nu e^{iu}\
,\ee
which can be written as
\be \int_{-1}^1 dt ~ g(t)~
\left[1+\frac{i}4(1-t)u\right]~e^{\frac{i}2(1-t)u}\ =\ 1-\nu
e^{iu}\ ,\ee
where $g=m\circ m$ with $\circ$ defined by
\cite{Prokushkin:1998bq}
 \be
 (p\circ q)(t)=\int_{-1}^1 ds \int_{-1}^1 ds'
 \delta(t-s s')~p(s)~q(s')\ .
 \ee
Replacing $iu$ by $-2d/dt$ acting on the exponential and
integrating by parts, we find
\be \int_{-1}^1 dt~\left[g(t)+\frac
12\left((1-t)g(t)\right)'\right]~e^{\frac{i}2(1-t)u}-\frac12\left[
(1-t)g(t)~e^{\frac{i}2(1-t)u}\right]_{-1}^1\ =\ 1-\nu e^{iu}\ .\ee
This can be satisfied by taking $g$ to obey
\be g(t)+\frac 12\left((1-t)g(t)\right)'\ =\ \d(1-t)\ ,\qquad
g(-1)\ =\ -\nu\ ,\ee
with the solution
\be (m \circ m)(t)\ =\ g(t)\ =\ \delta(t-1)-\frac{\nu}2 (1-t)\
.\ee

Even and odd functions are orthogonal with respect to the $\circ$
product, \emph{i.e.}
\be p^{(\s)}\circ q^{(\s')}\ =\ \d_{\s\s'} p^{(\s)}\circ
q^{(\s')}\ ,\qquad p^{(\s)}(-t)\ = \s p^{(\s)}(t)\ ,\quad \s=\pm 1
\ .\ee
Therefore
\bea (m^{(+)}\circ m^{(+)})(t)&=&I^{(+)}_0(t)-\frac{\nu}2\ ,\label{mmm}\\[5pt]
(m^{(-)}\circ m^{(-)})(t)&=&I^{(-)}_0(t)+\frac{\nu}2 t\
,\label{mmp}\eea
where
\be I^{(\pm)}_0(t)\ =\
\frac12\left[\delta(1-t)\pm\delta(1+t)\right]\ .\ee

One proceeds \cite{Prokushkin:1998bq}, by expanding $m^{(\pm)}(t)$
in terms of $I^{(\pm)}_0(t)$ and the functions ($k\geq 1$)
 \bea
 I^{(\s)}_k(t)&=&\left[{\rm sign}(t)\right]^{\frac12(1-\s
 )}~\int_{-1}^1 ds_1 \cdots \int_{-1}^1 ds_k~\delta(t-s_1\cdots
 s_k)\nn\\[5pt]
 &=&\left[{\rm sign}(t)\right]^{\frac12(1-\s)}{\left(\log
 \frac1{t^2}\right)^{k-1}\over (k-1)!}\ ,
 \eea
which obey the algebra ($k,l\geq 0$)
 \be
 I^{(\s)}_k\circ I^{(\s)}_l\ =\ I^{(\s)}_{k+l}\ .
 \label{ring}
 \ee
Thus, given a quantity
 \be
 p^{(\s)}(t)\ =\ \sum_{k=0}^\infty p_k I^{(\s)}_k(t)\ ,
 \ee
and defining its symbol
 \be
 \widetilde p^{(\s)}(\xi)\ =\ \sum_{k=0}^\infty p_k ~\xi^k\ ,
 \ee
it follows from \eq{ring} that
 \be
 (\widetilde{p^{\s}\circ q^{\s}}) (\xi)= \widetilde p^{(\s)} (\xi) \widetilde q^{(\s)}(\xi)\
 ,
 \ee
so that \eq{mmm} and \eq{mmp} become the algebraic equations
 \bea
 (\widetilde m^{(+)}(\xi))^2&=&1 - \frac{\nu}{2}\xi\ ,\nn\\[5pt]
 (\widetilde m^{(-)}(\xi))^2&=&1 + \frac{\nu}{2}{\xi\over 1+\frac12 \xi}\ .
\eea
Therefore
 \bea
 m^{(+)}(t)&=&\pm\left[ I^{(+)}(t) + q^{(+)}(t)\right]\
 ,\qquad \widetilde q^{(+)}(\xi)\ =\ \sqrt{1 - \frac{\nu}{2}\xi}-1\ ,\\[8pt]
 m^{(-)}(t)&=&
 \pm \left[I^{(-)}(t)+q^{(-)}(t)\right]\ ,\qquad \widetilde q^{(-)}(\xi)\
 =\ \sqrt{1 + \frac{\nu}{2}{\xi\over 1+\frac12 \xi}}-1\ .\nn
 \eea
The physical gauge condition \eq{physgauge}, which requires
 \be
 m(t)|_{\nu=0}\ =\ \delta(1+t)\ ,
 \ee
implies that
 \bea
 m(t)&=& m^{(+)}(t)-m^{(-)}(t)\ =\ \delta(1+t)+q(t)\ ,\w2 q(t)&=&
 q^{(+)}(t)-q^{(-)}(t)\
 .\label{meq}
 \eea

To obtain the functions $q^{(\pm)}(t)$ explicitly, we first expand
\be q^{(\pm)}(t)\ =\ \left[{\rm sign}(t)\right]^{\frac12(1-\s
 )} \sum_{k=1}^\infty \widetilde q^{(\pm)}_k {\left(\log
{1\over t^2}\right)^{k-1}\over (k-1)!}\ ,\ee
where the coefficients are related to the expansions of the
symbols as
\be \widetilde q^{(\pm)}(\xi)\ =\ \sum_{k=1}^\infty \widetilde
q^{(\pm)}_k \xi^k\ .\ee
In the case of $q^{(+)}(t)$, expansion of $\sqrt{1 -
\frac{\nu}{2}\xi}-1$ yields
\be q^{(+)}(t)\ =\ \sum_{k=0}^\infty {\frac 12 \choose
k+1}\left(-{\nu\over 2}\right)^{k+1}{\left(\log{1\over
t^2}\right)^{k}\over k!}\ =\ -{\nu\over 4}~{}_1\!
F_1\left[\frac12;2;{\nu\over 2}\log \frac 1{t^2}\right]\
.\label{qplus}\ee

In the case of $q^{(-)}(t)$, we begin by defining
\be q^{(-)}(t)\ =\ {\rm sign}(t) ~\widehat q(\log {1\over t^2})\
.\ee
Thus, from
\be \widehat Q(\zeta)\ \equiv\ \int_0^\zeta d\zeta' \widehat
q(\zeta')\ =\ \sum_{k=1}^\infty \widetilde q^{(-)}_k {\zeta^k\over
k!}\ ,\ee
and $\xi^k=\int_0^\infty d\zeta e^{-\zeta} {(\xi\zeta)^k\over
k!}$, it follows that $\widetilde q^{(-)}(\xi)=\int_0^\infty
d\zeta e^{-\zeta} \widehat Q(\xi\zeta)$. This Laplace-type
transformation can be inverted as
\be \widehat Q(\zeta)\ =\ \int_{\c-i\infty}^{\c+i\infty} {dz
\,e^{\zeta z}\over 2\pi i z} \widetilde q^{(-)}({1\over z})\ ,\ee
with
$$\c >\max \left\{ {\rm Re} z_i~:\ \mbox{$z_i$ pole or branch
cut of ${1\over z}\widetilde q^{(-)}({1\over z})$}\right\}\ .$$
The function $q^{(-)}(t)$ is then obtained by differentiation with
respect to $\zeta$ and the substitution $\zeta=\log {1\over t^2}$,
that is
\be q^{(-)}(t)\ =\ {\rm
sign}(t)\left.\int_{\c-i\infty}^{\c+i\infty} {dz \,e^{\zeta
z}\over 2\pi i } \widetilde q^{(-)}({1\over
z})\right|_{\zeta=\log{1\over t^2}}\ .\ee
The contour can be closed around the branch-cut that goes from
$z=-\frac12$ to $z=-\frac{1+\nu}2$, and one finds
\be q^{(-)}(t)\ =\ {\nu t\over 4}~{}_1\!F_1\left[\frac
12;2;-{\nu\over 2}\log{1\over t^2}\right]\ .\label{qminus}\ee

In summary, the internal solution is given by
 \bea
 \widehat \Phi'&=&{\nu\over b_1}\ ,\label{intsol1}\\[4pt]
 \widehat A'_\a&=& \frac{i}2 z_\a \int_{-1}^1
 dt~ q(t)\,
 e^{\frac{i}2(1+t)u}\ ,
 \\[4pt] q(t)&=&-{\nu\over 4}\left({}_1\!
F_1\left[\frac12;2;{\nu\over 2}\log \frac 1{t^2}\right]+t\,{}_1\!
F_1\left[\frac12;2;-{\nu\over 2}\log \frac 1{t^2}\right]\right)\ .
 \label{intconn}
 \eea

Expanding $\exp ({itu\over 2})$ results in integrals of the
degenerate hypergeometric functions times $t^p$ ($p=0,1,\dots$),
which improve the convergence at $t=0$. Thus $\widehat A_\a$ is a
formal power-series expansion in $u$ with coefficients that are
functions of $\nu$ that are well-behaved provided this is the case
for the coefficient of $u^0$. This is the case for $\nu$ in some
finite region around $\nu=0$, as we shall see next.


\scss{The Solution in Spacetime}\label{sec:spacetime}


Let us evaluate the physical fields in the two regions $\l^2x^2<1$
and $\l^2\widetilde x^2<1$ using the two gauge functions \eq{wL}
and \eq{wtL}. We first compute $\widehat \Phi=L^{-1}\star \widehat
\Phi'\star \pi(L)={\nu\over b_1} L^{-1}\star L^{-1}$, with the
result
\be \widehat \Phi\ =\ {\nu\over b_1}
(1-\lambda^2x^2)\exp\left[-i\l x^{\a\ad}y_\a \yb_{\ad}\right]\
.\label{wphi}\ee
This shows that the physical scalar field is given in the
$x^a$-coordinate chart by
\be \phi(x)\ =\ \widehat \Phi|_{Y=Z=0}\ =\ {\nu\over b_1}
(1-\lambda^2x^2)\ ,\qquad \l^2x^2<1\ ,\label{scalarsol}\ee
while the Weyl tensors for spin $s=2,4,\dots$ vanish. Using
instead $\widetilde L$, the physical scalar field in the
$\widetilde x^a$-coordinate chart is given by
\be \widetilde \phi\ =\ \nu(1-\lambda^2\widetilde x^2)\ ,\qquad
\l^2\widetilde x^2<1\ .\ee
As a result, the two scalar fields are related by a duality
transformation in the overlap region
\be \widetilde\phi(\tilde x)\ =\ {\nu\phi(x)\over \phi(x)-\nu}\
,\qquad \lambda^2 x^2=(\lambda^2\widetilde x^2)^{-1}<0\
.\label{duality}\ee
Thus, if the transition takes place at $\lambda^2
x^2=\lambda^2\widetilde x^2=-1$, then the amplitude of the
physical scalar never exceeds $2\nu$ so the open-string theory of
\cite{Johan}, which couples to the Weyl zero-form, is weakly
coupled throughout the solution. We also note that a
gauge-invariant characterization of the solution is provided by
the invariants \eq{Cninv}, that are given by

\be {\cal C}^-_{2p}\ =\ (\nu/b_1)^{2p}\ ,\label{onshellcharge}\ee
while ${\cal C}^+_{2p}$ diverges.

The gauge fields are defined by the decomposition \eq{Amu} with
$\widehat A_\mu=L^{-1}\star\partial_\mu
L=e_\mu^{(0)}+\omega_\mu^{(0)}$, that is
\be e_\mu+W_\mu\ =\ e_\mu^{(0)}+\omega_\mu^{(0)}-K_\m\
,\label{Asol}\ee
with $e_\mu^{(0)}$ and $\omega_\mu^{(0)}$ denoting the $AdS_4$
vacuum, and
\be K_\m\ =\ i \o_\m{}^{\a\b} L^{-1}\star \widehat A'_\a\star
\widehat A'_\b\star L-{\rm h.c.}\ ,\ee
which can be rewritten using \eq{zz} as
\bea K_\m&=&{1\over 16i}\o_\m{}^{\a\b} \int_{-1}^1 dt\int_{-1}^1
dt'~
q(t)q(t')\times\\
&&\times
\left.\left\{\left[(1+t)(1+t'){\partial^2\over\partial\r^\a\partial\r^\b}-
(1-t)(1-t'){\partial^2\over\partial\t^\a\partial\t^\b}\right]
V(y,\yb;\r,\t;tt')\right\}\right|_{\r=\t=0}\nn\\&&-{\rm h.c.}\
,\nn\eea
where
\be V(y,\yb;\r,\t;tt')\ =\ \left. \left[L^{-1}\star
e^{i\left(\frac{1-tt'}2 u+\r y+\t z\right)}\star
L\right]\right|_{Z=0}\ .\ee
Using \eq{gauss3}, this quantity can be expressed as
\be V(y,\yb;\r,\t;tt')\ =\ {4h^2\over (1+h)^2(1-tt' a^2)^2}\exp\,
i\,{\r((1+a^2)y+2a\yb)-2a^2\r\t\over 1-tt'a^2}\ ,\ee
where
\be  a_{\a\ad}\ =\ {\l x_{\a\ad}\over 1+\sqrt{1-\l^2 x^2}}\
.\label{defa}\ee
We can thus write
\be K_\m\ =\ {1\over 4i} Q \o_\m{}^{\a\b}\left[(1+a^2)^2y_\a
y_\b+4(1+a^2)a_\a{}^{\ad}y_\b\yb_{\ad}+4a_\a{}^{\ad}
a_\b{}^{\bd}\yb_{\ad}\yb_{\bd}\right]-{\rm h.c.}\ ,\label{Kmu}\ee
where
\be Q\ =\ -\frac14 (1-a^2)^2\int_{-1}^1 dt\int_{-1}^1 dt'~ {
q(t)q(t')(1+t)(1+t')\over (1-tt' a^2)^4}\ .\label{Q}\ee
This function is studied further and evaluated at order $\nu^2$ in
Appendix \ref{app:Q}. From \eq{Asol} and \eq{Kmu} it follows that
all higher-spin gauge fields vanish,
\be W_\m\ =\ 0\ ,\ee
while the vierbein and Lorentz connection are given by
\bea \o_{\m}{}^{\a\b}&=& f \o_{\m(0)}{}^{\a\b}\ ,\label{sol:lorconn}\\[8pt]
e_{\m}{}^{\a\ad}&=& e_{\m(0)}{}^{\a\ad}-\left(Q f+\bar Q\bar
f\right)\left[4a^2 e_{\m(0)}{}^{\a\ad} +{(1+a^2)^4\over
(1-a^2)^2}\l^3 dx^a x_a x^{\a\ad}\right]\ ,\label{sol:vierbein}\eea
where we have used $a_\a{}^{\ad}
a_\b{}^{\bd}\bar\o^{(0)}_{\ad\bd}=a^2\o^{(0)}_{\a\b}$, and defined
\be f \ =\ {1+(1-a^2)^2\bar Q\phantom{{\hat f}\over {\hat f}}\over
|1+(1+a^2)^2Q|^2-16a^4|Q|^2\phantom{{\hat f}\over a}}\
.\label{f}\ee
We identify the vielbein as a conformally rescaled AdS$_4$ metric,
\be e^a\ =\ f_1 dx^a + \l^2 f_2 dx^b  x_b x^a\ =\ {2 \O d(g_1
x^a)\over 1-\l^2 g_1^2 x^2}\ ,\ee
where
\be f_1(x^2)\ =\ {2(1+a^2)^2\over(1-a^2)^2}\Big(1-4a^2(Q f+\bar
Q\bar f)\Big)\ ,\qquad f_2(x^2)\ =\
{(1+a^2)^4\over(1-a^2)^2}\Big(Q f+\bar Q\bar f\Big)\ ,\ee
and the scale factor is given by
\be \O\ =\ {(1-\l^2 g^2_1 x^2 )f_1\over 2g_1}\ ,\qquad g_1\ =\
\exp \left[-\l^2 \int^{\l^{-2}}_{x^2}{f_2(t) dt\over
f_1(t)}\right]\ . \label{scale}\ee
In the boundary region, where $a^2\rightarrow 1$, the double
integral in \eq{Q} diverges at $t=t'=\pm 1$ while the pre-factor
goes to zero, as to produce a finite residue given by
\be \lim_{a^2\ra 1} Q\ =\ -{\nu^2\over 6}\ .\ee
In this limit
\be \lim_{a^2\ra 1}\O\ =\ {1\over 1-{4\nu^2\over 3}}\ .\ee
This factor is positive in the Type B model, while curiously
enough it blows up at a critical value, within the range
\eq{rangle}, in the Type A model.

By the $Z_2$-symmetry, the metric in the $\widetilde
x^a$-coordinate chart is given by the same functions as in the
$x^a$-coordinate system (which is not the same as a
reparameterization!). In the distant past and future, where
$a^2\rightarrow -1$, the function $Q$ diverges logarithmically
leading to qualitatively different behavior of the scale factors
in the cases of the Type A and B models, which is analyzed in more
detail in \cite{Varna}. Here we content ourselves by observing
that any pathological behavior in the strong-coupling region
$\lambda^2 x^2<-1$ can be removed by going to the dual weakly
coupled frame. Thus, geometrically speaking, the solution
interpolates between two asymptotically $AdS_4$ regions at
$\lambda^2x^2\sim 1$ and $\lambda^2\widetilde x^2\sim 1$ via an
interior given by complicated scale factors (discussed in
\cite{Varna}) times foliates determined by symmetries, to which we
shall turn our attention next.


\scss{Symmetries of the Solution}


A more general discussion of symmetries of solutions will be given
in Section \ref{sec:5}. In view of \eq{epsprime}, the gauge
transformations preserving the primed solution obey
\be \widehat D'_\a\widehat\e'\ =\ 0\ ,\qquad \pi(\widehat
\epsilon')\ =\ \widehat \e'\ ,\label{symm1}\ee
where the last condition is equivalent to
$[\widehat\epsilon',\widehat \Phi']_\pi=0$ since
$\widehat\Phi'=\nu$ is constant. The condition \eq{symm1} is by
construction solved by the full $SO(3,1)$ generators, \emph{i.e.}
\be \widehat\epsilon^\prime\ =\ \frac1{4i}\Lambda^{\a\b}\widehat
M'_{\a\b}-{\rm h.c.}\ ,\ee
with $\widehat M'_{\a\b}$ given by \eq{Mhat} and constant
$\Lambda_{\a\b}$.

The solution is also left invariant by additional transformations
with rigid higher-spin parameters
\be \widehat \e'\ =\ \sum_{\ell=0}^\infty \widehat\e'_{\ell}\
,\label{hsl21}\ee
where the $\ell$'th level is given by
\be \widehat \e'_\ell\ =\ \sum_{m+n=2\ell+1} \Lambda^{\a_1\dots
\a_{2m},\ad_1\dots \ad_{2n}}\widehat M'_{\a_1\a_2}\star\cdots
\star \widehat M'_{\a_{2m-1}\a_{2m}}\star\widehat
M'_{\ad_1\ad_2}\star\cdots\star\widehat
M'_{\ad_{2n-1}\ad_{2n}}-{\rm h.c.}\ ,\label{hsl22}\ee
with constant $\Lambda^{\a_1\dots\a_{2m},\ad_1\dots\ad_{2n}}$.
These parameters span the solution space to \eq{symm1}, provided
that this space has a smooth dependence on $\nu$. The full
symmetry algebra is thus a higher-spin extension of $SO(3,1)\simeq
SL(2,C)$, that we shall denote by
\be hsl(2,C;\nu)\supset sl(2,C)\ ,\ee
where $sl(2,C)$ is generated by $\widehat M'_{\a\b}$ and its
hermitian conjugate, and we have indicated that in general the
structure coefficients may depend on the deformation parameter
$\nu$.

The generators of the $SO(3,1)$ transformations preserving the
space-time dependent field configuration, that is, obeying
\eq{eps}, are by construction given by
\be \widehat M^L_{\a\b}\ =\ L^{-1}(x) \star\widehat M'_{\a\b}\star
L(x)\ ,\ee
and are related to the full Lorentz generators $\widehat M_{\a\b}$
by
\be \widehat M^L_{\a\b}-\widehat M_{\a\b}\ =\ \widetilde
M_{\a\b}(\l x) -M_{\a\b}\ ,\ee
where $M_{\a\b}=y_\a y_\b$ and we have defined
\be \widetilde M_{\a\b}(v)\ =\ L^{-1}(\lambda^{-1}v)\star
M_{\a\b}\star L(\lambda^{-1}v)\ =\ \widetilde y_\a(v)\widetilde
y_\b(v)\ ,\label{Mtilde}\ee
where the transformed oscillators are defined by
\be \widetilde y_\a(v)\ =\ {y_\a+v_\a{}^{\ad}\yb_{\ad}\over
\sqrt{1-v^2}}\ ,\label{ytilde}\ee
and obey the same algebra as the original oscillators, \emph{viz.}
\be \widetilde y_\a\star\widetilde y_\b\ =\ \widetilde
y_\a\widetilde y_\b + i\e_{\a\b}\ .\ee

If we let $\widehat g'_\L$ and $\widehat g^L_\L$ be the group
elements generated by $\widehat M'_{\a\b}$ and $\widehat
M^L_{\a\b}$, respectively, then it follows that
\be L(x)\star \widehat g_{\Lambda}\ =\ \widehat g'_\Lambda\star
L(\Lambda x)\ ,\qquad (\Lambda x^\mu)=\Lambda^\mu{}_\nu x^\nu\
.\label{lor}\ee
The spacetime decomposes under this $SO(3,1)$ action into orbits
that are three-dimensional hyper-surfaces which describe local
foliations of $AdS_4$ with $dS_3$ and $H_3$ spaces in the regions
$x^2>0$ and $x^2<0$, respectively.


\scs{On Other Solutions With Non-Maximal Isometry}\label{sec:5}


In this section we discuss the consequences of imposing various
non-maximal symmetry conditions on solutions. We first do this in
a general setting, and then consider symmetry groups of dimensions
3, 4 and 6. In this approach, we recover the previous
$so(3,1)$-invariant solution and also construct new solutions at
the first order in the Weyl zero-form. The latter include domain
walls and rotationally invariant solutions.

\scss{Some Generalities}

The solution with maximal unbroken symmetry is the AdS$_4$ vacuum
$\widehat \Phi=0$, which is invariant under rigid $hs(4)$
transformations, with $Z$-independent parameters. Let us consider
non-vanishing $\widehat\Phi$ that is invariant under a non-trivial
set of transformations with parameters belonging to
\be h(\widehat\Phi)\ =\ \left\{\widehat\e\ :\quad \widehat
D\widehat \e\ =\ 0\ ,\quad[\widehat\e,\widehat\Phi]_\pi\ =\ 0\
,\quad \t(\widehat\e)=\widehat\e^\dagger=-\widehat\e\right\}\
.\label{eps}\ee
As is the case for $hs(4)$, this algebra closes under
$\star$-commutation of parameters as well as compositions induced
by the associativity of the $\star$-product, \emph{e.g.}
\be
\widehat\e_1\star\widehat\e_2\star\widehat\e_3+\widehat\e_3\star\widehat\e_2\star\widehat\e_1\
,\qquad \widehat\e_i\in h(\widehat\Phi)\ .\ee
In the case that $h(\widehat\Phi)$ contains a finite-dimensional
rank-$r$ subalgebra $\mathbf g_r\subset SO(3,2)$ this induces a
natural higher-spin structure $h(\widehat\Phi)$, as exemplified in
\eq{hsl21} and \eq{hsl22} for $hsl(2,C;\nu)$.

In the $Z$-space approach
\be \widehat\e\ =\ \widehat L^{-1}\star\widehat \e'\star \widehat
L\ ,\qquad \partial_\m\widehat\e'\ =\ 0\ ,\ee
where
\be \widehat D'_\a\widehat\e'\ =\ 0\ ,\qquad
[\widehat\e',\widehat\Phi']_\pi\ =\ 0\ .\label{epsprime}\ee
Expanding perturbatively in $\widehat\Phi'|_{Z=0}=C'$,
\be \widehat \e'\ =\ \e'+\widehat \e_{(1)}'+\widehat
\e_{(2)}'+\cdots\ ,\label{epsexp}\ee
and assuming that the parameters obeying \eq{epsprime} are given
up to and including order $n-1$ ($n=1,2,\dots$), then
$\widehat\e'_{(n)}$ is determined by
\be (\widehat D'_\a\widehat\e')_{(n)}\ =\ 0\ ,\ee
which is an integrable partial differential equation in $Z$-space
provided that
\be \widehat I'_{(n)}\ \equiv
([\widehat\e',\widehat\Phi']_\pi)_{(n)}\ =\ 0\ .\label{intcond}\ee
Using the $Z$-space field equations obeyed in the lower orders,
one can show that
\be \partial_\a \widehat I'_{(n)}\ =\ 0\ ,\ee
so that \eq{intcond} holds if ($n=1,2,\dots$)
 \be I'_{(n)}\ \equiv\ \left.\widehat I'_{(n)}\right|_{Z=0}\ =\
\sum_{p+q=n}\left.[\widehat \e'_{(p)},\widehat
\Phi'_{(q)}]_\pi\right|_{Z=0}\ =\ 0\ .\label{iprn}\ee

In the first order, the symmetry condition reads
\be [\e',C']_\pi\ =\ 0\ .\ee
We shall denote the stability algebra of $C'$
\be h(C')\ =\ \left\{\ \e'\ :\quad [\e',C']_\pi\ =\ 0\ ,\qquad
\t(\e')\ =\ (\e')^\dagger\ =\ -\e'\ \right\}\ .\ee
As found in the case of $hsl(2,C;\nu)$, the full symmetry algebra
$h(\widehat\Phi)$ is in general a deformed version of $h(C')$.
Moreover, we shall denote the space of all twisted-adjoint
elements invariant under $h(C')$ by $B(C')$, \emph{i.e.}
\be B(C')\ =\ \left\{\ \widetilde C'\ :\quad [\e',\widetilde
C']_\pi=0\ ,\quad \forall\e'\in h(C')\ \right\}\ .\label{bcpr}\ee
Covariance implies that if $g$ is an $hs(4)$ group element, then
\be h\left(g^{-1}\star C'\star\pi(g)\right)\ =\ g^{-1}\star
h(C')\star g\ .\ee
The spaces $B(g^{-1}\star C'\star\pi(g))$ and $g^{-1}\star
B(C')\star \pi(g)$ are in general not isomorphic, however, as
there exist special points in the twisted-adjoint representation
space where $\dim B(C')$ is less than the generic value on the
group orbit. The simplest example is the point $C'=\nu$, as will
be discussed below \eq{rhopm}. Another subtlety, that we shall
exemplify below, is related to the fact that the associativity of
the $\star$-product implies that if $C'$ is a regular initial
condition then $B(C')$ contains the elements $C'_{(2n+1)}$ defined
by \eq{c2n1}. Thus, if $\dim B(C')$ is finite then $C'$ must
either violate regularity at some level of perturbation theory or
be projector-like in the sense that $C'_{2n+1}\sim C'$ for some
finite value of $n$.

In the case of a solution in which $\widehat\Phi$ asymptotes to
$L^{-1}\star C'\star L$ with $h(C')\supset \mathbf g_r\subset
SO(3,2)$, where \eq{l} is the AdS$_4$ gauge function \eq{l}, the
solution has $\mathbf g_r$ isometry close to the boundary provided
the perturbation theory holds. Since the space-time field
equations are manifestly diffeomorphism and locally Lorentz
invariant, the $\mathbf g_r$-isometry extends to the solution in
the interior, where it acts on the full master fields via
parameters $\widehat\e=L^{-1}\star\widehat\e'\star L$. Hence, the
integrability conditions \eq{iprn} must hold, resulting in a full
symmetry algebra $h(\widehat\Phi)$ that is in general some
deformed higher-spin extension of $\mathbf g_r$ with deformation
parameters given by $C'$.

Let us next examine the above features in more detail in some
special cases.


\scss{Solutions with Unbroken $SO(3,1)$ Symmetry}


Acting on the exact $SO(3,1)$-invariant solution described in
Section \ref{sec:4} with the gauge transformation generated by the
group element
\be g(v)\ =\ L(\l^{-1}v)\ ,\ee
with $L$ given by \eq{wL}, which requires
\be v^2\ <\ 1\ ,\ee
one finds the gauge-equivalent exact solution given by the
zero-form
\be \widehat \Phi^{\prime(v)}\ =\ g^{-1}(v)\star \widehat
\Phi'\star \pi(g(v))\ =\ {\nu\over b_1} (1-v^2) \exp (i y v\yb)\
,\label{new}\ee
and the internal connection
\be \widehat A'^{(v)}_\a\ =\ {i\over 2}\int_{-1}^1
dt\left[q^{(+)}(t)-q^{(-)}(t)\right] g^{-1}(v)\star(z_\a e^{\frac
i2(1+t)u})\star g(v)\ .\label{Av}\ee

The gauge-transformed solution has
\be C^{\prime(v)}\ =\ \widehat \Phi^{\prime(v)}\ ,\ee
with stability group $h(C^{\prime(v)})$ generated via enveloping
of the generators $\widetilde M_{\a\b}(v)$ given by \eq{Mtilde}.
Thus, the space $B(C^{\prime(v)})$ consists of all elements
obeying
\be \left[\,\widetilde y_\a(v)\widetilde y_\b(v)\,,\,\widetilde
C'\,\right]_\pi\ =\ 0\ ,\label{vso31}\ee
amounting to the following second order partial differential
equation
\be {4\over 1-v^2}\Big(i y_\a\partial_\b +i (v\yb)_\a (v\bar
\partial)_\b+y_\a (v \yb)_\b
-\partial_\a (v\bar\partial)_{\b}\Big)\widetilde
C'+(\a\leftrightarrow\b)\ =\ 0\ . \ee
This equation can be solved using the ansatz
\be \widetilde C'\ =\ \widetilde C'(V)\ ,\qquad V\ =\ yv\yb\ ,\ee
implying the following second-order ordinary differential equation
in variable $V$ with constant coefficients
\be \left(-v^2{d^2\over dV^2}+i(1+v^2){d\over
dV}+1\right)\widetilde C'\ =\ 0\ ,\label{ddV}\ee
which admits the solutions
\be \widetilde C'\ =\ \mx{\{}{ll}{\widetilde \nu_1 e^{i
V}+\widetilde\nu_2 e^{i{V\over
v^2}}&\mbox{for $v^2\neq 0$ and $v^2<1$}\\[5pt] \widetilde\nu_1
e^{i V}&\mbox{for $v^2=0$}}{.}\ .\label{cprv}\ee
It is not possible to produce any further solutions to \eq{vso31}
using \eq{c2n1}, since $\exp{iV}\star\pi(\exp{iV})$ is
proportional to $1$, while the $\star$-product composition of
$\exp{iV}$ and $ \pi(\exp{i{V\over v^2}})$ is divergent. Thus,
\be \dim B(C^{\prime(v)})\ =\ \mx{\{}{ll}{2&\mbox{for $v^2\neq 0$ and $v^2<1$}\\[5pt]
1&\mbox{for $v^2=0$}}{.}\ . \ee
We stress that $h(C^{\prime(v)})$ is the stability group of the
twisted-adjoint element $C^{\prime(v)}$, and not of the parameter
$v$, and that therefore the stability group is still $SO(3,1)$
when $v$ is null. The additional solution with super-luminal
boost-parameter $v^a$ is not gauge-equivalent to $\widehat
\Phi^{\prime (v)}$, and that $\widetilde C'$ is regular or
singular as a initial condition, according to the terminology
introduced in Section \ref{sec:sing}, depending on whether
$\widetilde\nu_1\widetilde\nu_2=0$ or
$\widetilde\nu_1\widetilde\nu_2\neq 0$, respectively. Whether the
singular or super-luminal cases can be elevated to exact solutions
remains to be seen.

For $v^2\neq 0$ the space $B(C^{\prime(v)})$ in invariant under
$v^a\leftrightarrow v^a/v^2$. Extending \eq{cprv} to
\be v^2\ >\ 1\ ,\ee
gives an $SO(3,1)$-invariant two-dimensional solution space with
stability group generated by $\widetilde M_{\a\b}(v)$ given by
\eq{Mtilde}, where the fact that the denominator in \eq{ytilde} is
imaginary does not present an obstacle since the $sl(2,C)$-doublet
oscillators are complex.

The ``self-dual'' case $v^2=1$ requires a separate treatment. Here
$\eta_\a=y_\a+v_\a{}^{\ad}\yb_{\ad}$ is a commuting oscillator
giving rise to a three-dimensional translation generator
$p_a=(v^b(\s_{ab})^{\a\b}\eta_\a \eta_\b+{\rm h.c.})$ obeying
$[p_a,p_b]_\star=0$ and $v^ap_a=0$. The commuting oscillators obey
the reality condition $(\eta_\a)^\dagger=v_{\ad}{}^\a\eta_\a$, and
transform as doublets under the oscillator realization of the
$SL(2,R)\simeq SO(2,1)$ that leaves $v_a$ invariant. This leads to
an $ISO(2,1)$-invariant two-dimensional solution space $B(e^V)$ to
be described next at the level of the linearized field equations
(see eq. \eq{Ciso21}) together with some other interesting
reductions.


\scss{On Domain Walls, Rotationally Invariant and RW-like
Solutions}\label{sec:symm}


Here we shall discuss some classes of solutions of considerable
interest corresponding to the rank-$r$ subalgebras $\mathbf
g_r\subset SO(3,2)$ parameterized by
\bea \mathbf g_3&:& M_{ij}\ =\ L^a_i L^b_j M_{ab}\
,\label{g3}\\[5pt]
\mathbf g_4&:& M_{ij}\ ,\quad P\ =\ L^a P_a\ =\ \frac14
L^a(\s_a)^{\a\ad}y_\a \yb_{\ad}\ ,\label{g4}\\[5pt]
\mathbf g_6&:& M_{ij}\ ,\quad P_i\ =\ ( \a M_{ab}L^b+\b P_a)L_i^a\
,\label{g6}\eea
where $\a$ and $\b$ are real parameters and, in all cases,
$(L_i^a,L^a)$ is a representative of the coset $SO(3,1)/SO(3)$ or
$SO(3,1)/SO(2,1)$, obeying
\be L^a L_a\ =\ \e\ =\ \pm 1\ ,\qquad L_i^a L_a\ =\ 0\ ,\qquad
L_i^a L_{ja}\ =\ \eta_{ij}\ =\ {\rm diag}(+,+,-\e)\ .\ee
The $SO(3,2)$ algebra \eq{so32b} yields $[M,M]\sim M$.
Furthermore, for $\mathbf g_4$ one finds $[M,P]\sim 0$, while for
$\mathbf g_6$ one finds $[M,P]\sim P$, and
 \be
 [P_i,P_j]\ =\ i(\b^2-\e \a^2) M_{ij}\ .
 \ee
In summary, one has
 \bea
 \mathbf g_6 &=& \mx{\{}{ll}
 {SO(3,1)&\mbox{for $\a^2-\e\b^2>0$\ ,\quad \ $\e=\pm 1$}
 \\[5pt]
 ISO(2,1)&\mbox{for $\b^2-\a^2=0$\ ,\qquad $\e=+1$}
 \\[5pt]
 SO(2,2)&\mbox{for $\b^2-\a^2>0$\ ,\qquad $\e=+1$}}{.}
 \\[10pt]
 \mathbf g_4 &=& \mx{\{}{ll}
 {SO(3)\times SO(2)\ , &\ \e=-1
 \\[5pt]
 SO(2,1)\times SO(2)\ , &\  \e=+1}{.}\\[5pt]
 \mathbf g_3 &=& \mx{\{}{ll}{SO(3)\ , &\ \e=-1
 \\[5pt]
 SO(2,1)\ , &\  \e=+1}{.}
 \eea

Next we seek $\mathbf g_r$-invariant twisted-adjoint initial
conditions $C'(y,\bar y)=\widehat\Phi'|_{Z=0}$ that obey the
invariance condition \eq{iprn} in \emph{the first order},
\emph{viz.}
 \be
 [\epsilon',C']_\pi\ =\ 0\ ,\qquad \e'\in \mathbf g_r\ .
 \label{grinv}
 \ee
As discussed in Section 3.2, $C'$ provides an initial condition to
\eq{z1}-\eq{z3} giving rise to a full master field $\widehat\Phi'$
related to the space-time Weyl tensors through
$\widehat\Phi=L^{-1}\star \widehat\Phi'\star L^{-1}$, so that
$C=L^{-1}\star C'\star L^{-1}$ are the linearized Weyl tensors,
discussed in Section 3.1. We note that $C$ is invariant under
gauge transformations with parameters $L^{-1}\star \e'\star L$,
while $\widehat \Phi$ is invariant under gauge transformations
with deformed parameters $\widehat \e'$ provided that the
integrability conditions \eq{iprn} hold to all orders. In Section
5.4 we shall perform the symmetry analysis in the second order.

The condition \eq{grinv} decomposes into two irreducible
conditions
\be [M_{ij},C']_\star\ =\ 0\ ,\qquad \mx{\{}{ll}{[P_i,C']_\pi\ =\
0&\mbox{for $\mathbf g_6$}\\[5pt] \{P,C'\}_\star\ =\ 0&\mbox{for
$\mathbf g_4$}}{.}\ .\label{grinv2}\ee
The first condition can be shown to have the general solution
 \be
 \mathbf g_3\ :\quad C'\ =\ C'(P)\ ,\ee
where $C'(P)$ is a function that we shall assume is analytic at
the origin. To arrive at this conclusion one can use the fact that
the oscillator realization \eq{mab} implies
$M_{[ab}M_{c]d}=M_{[ab}P_{c]}=M_{ab}M^b{}_c=M_{ab}P^b=0$, or,
alternatively, note that the only $\mathbf g_3$-invariant
spinorial objects are $\e_{\a\b}$, $\e_{\ad\bd}$ and $L_{\a\ad}$.

Thus, the $\mathbf g_3$-invariant solution space $B(C')$, defined
by \eq{bcpr}, is infinite-dimensional, and indeed closes under
\eq{c2n1} for regular initial data. The $\mathbf g_3$-invariance
can be imposed at the full level using the deformed generators
\be \mathbf g_3\ :\quad \widehat M'_{ij}\ =\ L_i^a  L^b_j \widehat
M'_{ab}\ ,\label{g3hat}\ee
where $\widehat M_{ab}$ are obtained from \eq{Mhat}. This results
in consistent $SO(3)$-invariant or $SO(2,1)$-invariant
``mini-superspace'' truncations described by the master equations
\eq{z1}-\eq{z3} with reduced master fields
\be \widehat\Phi'\ =\ \widehat\Phi'(P;u,\bar u;P',\Pi,\bar\Pi)\
,\ee
and
\be \widehat A'_\a\ =\ z_\a \widehat A_1+y_\a \widehat
A_2+L_\a{}^{\ad} (\zb_{\ad} \widehat A_3+ \yb_{\ad} \widehat A_4)\
,\qquad \widehat A_{i}\ =\ \widehat A_i(P;u,\bar
u;P',\Pi,\bar\Pi)\ ,\ee
taken to depend on the oscillators only through the following set
of composite variables,
\be u\ =\ y^\a z_\a\ ,\quad P\ =\ \frac14 L^{\a\ad}y_\a \yb_{\ad}\
,\quad P'\ =\ \frac14 L^{\a\ad}z_\a \zb_{\ad}\ ,\quad \Pi\ =\
L^a(\s_a)^{\a\ad}z_\a \yb_{\ad}\ ,\ee
where $L_{\a\ad}=(\s^a)_{\a\ad}L_a$. We stress that the full
$\mathbf g_3$ symmetry is manifest, assuring that the above
restricted dependence on the oscillators is actually consistent
with the master equations \eq{z1}-\eq{z3}. The reduced models
correspond geometrically to local reductions on either a
two-sphere or a two-hyperboloid, with complete symmetry algebra
given by deformed enveloping-algebra extensions of $\mathbf g_3$
generated by $\widehat M'_{ij}$.

Turning to the second set of conditions in \eq{grinv2}, we use the
lemmas
\bea L_i{}^a L^p[M_{ab},P^k]_\star &=& ik\e L_i{}^a P_a P^{k-1}\
,\\[5pt] L_i{}^a \{P_a,P^k\}_\star&=& L_i{}^a P_a\left(
2P^k-{k(k-1)\e\over 8}P^{k-2}\right)\ ,\\[5pt] \{P,P^k\}_\star&=&
2P^{k+1}+{k(k+1)\e\over 8}P^{k-1}\ ,\eea
to rewrite them as the following second-order differential
equations
\bea {\mathbf g_6}&:& \left(-{\b \e\over 8}{d^2\over dP^2}+i\a
\e{d\over dP}+2\b\right) C'(P)\ =\ 0\ ,\\[5pt] {\mathbf g_4}&:&
\left({\e\over 8}{d^2\over dP^2}+2\right)P C'(P)\ =\ 0\ .\eea
The resulting initial data that are analytic at $P=0$ are given by
 \bea
 SO(3,1)&:&C'(P)\ =\ \nu_1 e^{4i \r_- P}+\nu_2 e^{4i\r_+P}\ , \label{Cso31}
 \\[5pt]
 ISO(2,1) &:& C'(P)\ =\ (\mu_1+i \mu_2 P)e^{4iP}\ ,\label{Ciso21}
 \\[5pt]
 SO(2,2)&:& C'(P)\ =\ (\nu_1+i\nu_2)
 e^{4i\r_-P}+(\nu_1-i\nu_2)e^{4i\r_+ P}\ ,
 \\[5pt]
 SO(3)\times SO(2)&:& C'(P)\ =\ \nu_1 {\sinh(4P)\over P}\ ,
 \label{Cso3}\\[5pt]
 SO(2,1)\times SO(2)&:& C'(P)\ =\ \nu_1 {\sin(4P)\over P}\
 ,\label{Cso21}
 \eea
where $\nu_{1,2}$ and $\mu_{1,2}$ are real constants and
\be
 \r_{\pm}\ = \frac{\a}{\b}\pm\sqrt{\left(\frac{\a}{\b}\right)^2-\e}\ .
\label{rhopm}\ee
We see that in general $\dim B(C')=2$ for the $\mathbf
g_6$-invariant initial data, except at a few special points where
$\dim B(C')=1$, while $\dim B(C')=1$ for the $\mathbf
g_4$-invariant initial data.

In the case of $SO(3,1)$, we can identify particular cases of
\eq{Cso31} with the exact solutions \eq{new} and \eq{Av},
namely\footnote{The case of $v^2=0$ requires the embedding of
$SO(3,1)$ into $SO(3,2)$ using \eq{Mtilde}.}
\bea \e\ =\ 1\ ,\quad 1>v^2>0&:&\quad \mx{\{}{ll}{L^a\ =\
{v^a\over \sqrt{v^2}}\ ,\quad \r_+\ =\ -\sqrt{v^2}\ ,& \nu_2\ =\
{\nu\over b_1}(1-v^2)\ ,\quad \nu_1=0\\[4pt]L^a\ =\ -{v^a\over \sqrt{v^2}}
\ ,\quad \r_-\ =\ \sqrt{v^2}\ ,& \nu_1\ =\ {\nu\over b_1}(1-v^2)\
,\quad\nu_2=0}{.}\ ,\nn\\[5pt] \e\ =\ -1\ ,\quad v^2<0 &:&\quad
\mx{\{}{ll}{L^a\ =\ -{v^a\over\sqrt{-v^2}}\ ,\quad  \r_+\ =\
\sqrt{-v^2}\ ,& \nu_2\ =\ {\nu\over b_1}(1-v^2)\ ,\quad \nu_1=0\\[4pt]
L^a\ =\ {v^a\over\sqrt{-v^2}}\ ,\quad\r_-\ =\ -\sqrt{-v^2}\ ,&
\nu_1\ =\ {\nu\over b_1}(1-v^2)\ ,\quad \nu_2=0}{.}\ .\nn\eea

On the other hand, as noted in Section 5.2, the initial condition
$C'=\tilde{\nu} e^{iyv\yb}$ with $v^2>1$ and $\tilde{\nu}$ an
arbitrary constant, and the initial conditions with
$\nu_1\nu_2\neq 0$ are gauge-inequivalent to the above exact
solutions. In these cases one might expect the full zero-form
$\widehat\Phi'$ to receive $Z$-dependent higher-order corrections.
In the perturbative approach, the case of $\nu_1\nu_2\neq 0$ is
singular in the sense defined in Section \ref{sec:sing}, while the
case of $v^2>1$ involves poles at $t=1/v^2<1$ that can be avoided
using suitable contours in the $t$-plane.

Turning to the $ISO(2,1)$-invariant case \eq{Ciso21}, we identify
it as a singular initial condition for a flat domain wall
solution. The case with $\mu_2=0$ can be obtained formally as the
limit of the $dS_3$ domain wall in which
\be v^2\rightarrow 1\ ,\qquad \left|{\nu\over
b_1}\right|\rightarrow \infty\ ,\qquad \mu_1\ =\ {\nu(1-v^2)\over
b_1}\ \mbox{fixed}\ .\ee
It remains to be seen whether the internal connection \eq{Av} and
the space-time gauge fields are well-behaved in this limit. In
view of \eq{rangle}, we expect there to be differences between the
limiting procedures in the Type A and the Type B model, where
$\nu$ is real and imaginary, respectively.

The remaining three types of initial conditions listed above,
\emph{i.e.} the those invariant under $SO(2,2)$, $SO(3)\times
SO(2)$ and $SO(2,1)\times SO(2)$, are singular for all values of
the parameters.

In the singular cases found above, the perturbative expansion can
be obtained using the closed contours given in \eq{oint1} and
\eq{oint2}. It remains to be seen, however, whether full solutions
can be given explicitly in these cases. Another issue to settle is
whether the $\e'$-parameters can be deformed into full
$\widehat\e'$-parameters obeying \eq{epsprime}, to which we turn
our attention next.


\scss{Existence of Symmetry Parameters to the Second Order}


Let us prove the existence of the second correction
$\widehat\e'_{(2)}$ to the $\widehat\e'$-parameter associated to
the $\mathbf g_r$-symmetries discussed above. To do so, we need to
establish the integrability of \eq{epsprime} to second order,
\emph{i.e.} verify \eq{iprn} for $n=2$, where the relevant
quantity reads
\be I'_{(2)}\ \equiv \
\left.\left([\widehat\e'_{(1)},C']_\pi+[\e',\widehat
\Phi'_{(2)}]_\pi\right)\right|_{Z=0}\ ,\label{Ipr2}\ee
with $\e'\in\mathbf g_r$ given in \eq{g6} or \eq{g4}, and $C'$ in
\eq{Cso31}-\eq{Cso21}.

In fact, it is possible to show that \eq{Ipr2} vanishes for the
more general case with parameter
\be \e'\ =\ \y'-\y'^\dagger\ ,\qquad \y'\ =\ \frac 1{4i}(\l^{\a\b}
y_{\a}y_{\b}+\l^{\a\ad}y_\a \yb_{\ad})\ ,\ee
and twisted-adjoint element
\be C'\ =\ \sum_{m=0}^\infty i^m f_m
v_{\a(m)\ad(m)}y^{\a(m)}\yb^{\ad(m)}\ ,\ee
where $f_m$ are real numbers; we use the notation
$y^{\a(m)}=y^{\a_1}\cdots y^{\a_m}$; and we have defined
\be v_{\a(m)\ad(m)}\ =\ v^{a_1}\cdots
v^{a_m}(\s_{a_1})_{\a_1\ad_1}\cdots (\s_{a_m})_{\a_m\ad_m}\
.\label{multiv}\ee
The first-order correction to the parameter is given by
\be \widehat\e_{(1)}\ =\
\widehat\eta_{(1)}-\widehat\eta^\dagger_{(1)}\ ,\ee
with
\be \widehat \eta_{(1)}\ =\ -b_1 \int_0^1 dt \int_0^1 t'dt'\left(
{itt'\over 2}\l^{\a\b} z_\a z_\b+
\l^{\a\bd}z_\a\bar\partial_{\bd}\right)C'(-tt'z,\yb) e^{itt'yz}\
,\ee
and the second-order correction to the zero-form is given in
\eq{Phi2}, which can be re-written as
\be \widehat\Phi'_{(2)}\ =\ T\widehat B\ ,\ee
where the twisted-adjoint projection map $T$ is defined by
\be T\widehat f\ =\ \widehat f+\t\bar\pi\widehat
f+\pi\left(\widehat f+\t\bar\pi\widehat f\right)^\dagger\ ,\ee
and
\be \widehat B\ =\ -z^\a\int_0^1 dt\left(\widehat A_\a^{(1)}\star
C'\right)_{Z\ra tZ}\ ,\ee
with $\widehat A_\a^{(1)}$ given by \eq{a1}. The quantity
$I'_{(2)}$, which is a twisted-adjoint element, can thus be
written as
\be I'_{(2)}\ =\ \left.T\left(\widehat B_1+\widehat
B_2\right)\right|_{Z=0}\ ,\ee
where
\be \widehat B_1\ =\ \widehat\eta'_{(1)}\star C'\ ,\ee
and
\be \widehat B_2\ =\ T[\e',\widehat B]_\pi\ .\ee
The expansion of $\widehat B_1$ reads
\bea \widehat B_1&=& -b_1\int_0^1 dt\int_0^1
t'dt'\sum_{m,n}(-tt')^m i^{m+n}f_m f_n v^{\a(m)\ad(m)}
v^{\b(n)\bd(n)}\times\nn\\&& \times\Bigg(\left({itt'\over
2}\l^{\c\d}z_\c
z_\d+\l^{\c\dd}z_\c\bar\partial_{\dd}\right)z_{\a(m)}\yb_{\ad(m)}
e^{itt'yz}\Bigg)\star y_{\b(n)}\yb_{\bd(n)}\ .\eea
To calculate the $\l^{\c\d}$-contributions to $T\widehat
B_1|_{Z=0}$ we let $n\rightarrow m+2+k$ and contract $m+2$ of the
$y_\b$-oscillators with the $z$-oscillators in the first factor.
The remaining $y_\b$-oscillators may contract $z$-oscillators in
the exponent, so we sum over $l$ such contractions with $0\leq
l\leq k$. The resulting terms have the common $l$-independent
structure
\bea && T\left(\l^{\c\d}
v^{\a(m)\ad(m)}v_{\a(m)\b(k)\c\d\bd(m+2+k)} y^{\b(k)}
~(\yb_{\ad(m)}\star \yb^{\bd(m+2+k)})\right)\nn\\&& \sim
T\left(\l^{\c\d} v_{\c\d\a(k)\ad(k+2)}
y^{\a(k)}\yb^{\ad(k+2)}\right)\ =\ 0\ ,\eea
where we use \eq{multiv} and the fact that
$y^{\a(k)}\yb^{\ad(k+2)}$ cannot contribute to the twisted-adjoint
representation. Similarly, to calculate the
$\l^{\c\dd}$-contributions we let $n\rightarrow m+1+k$, resulting
in
\bea &&T\left(\l^\c{}_{\ad}
v^{\a(m)\ad(m)}v_{\c\a(m)\b(k)\bd(m+1+k)}y^{\b(k)}(\yb_{\ad(m-1)}\star
\yb^{\bd(m+1+k)})\right)\nn\\ && \sim T\left(\l^\c{}_{\ad}
v_{\c\b(k)\bd(k+1)} y^{\b(k)} \yb^{\ad\bd(k+1)}\right)\ =\ 0\
.\eea
The same type of cancellations occur in $T\widehat B_2|_{Z=0}$.
Here the $\l^{\c\d}$-contribution to $[\e',\widehat B]_\pi$ is
given by
\bea&& {b_1\over 8} \int_0^1 dt\int_0^1 t'dt' \sum_{m,n}(-t')^m
i^{m+n}f_m f_n v^{\a(m)\ad(m)} v^{\b(n)\bd(n)}\times\nn\\&&
\times~ \l^{\c\d}\left[y_\c
y_\d\,,\,z^\a\left((z_{\a(m+1)}\yb_{\ad(m)}e^{it'yz})\star
(y_{\b(n)}\yb_{\bd(n)})\right)_{Z\ra tZ}\right]_\star\ ,\eea
which we evaluate at $Z=0$ using
\be\l^{\c\d}[y_\c y_\d,z^\a\widehat f]_\star|_{Z=0}\ =\
4\l^{\c\a}\partial^{(y)}_\c\widehat f|_{Z=0}\ ,\ee
resulting in contributions to $T[\e',\widehat B]_\pi|_{Z=0}$ of
the form
\bea&& T\left(\l^{\a\b}v^{\a(m)\ad(m)} v_{\a(m+1)\b(k)\bd(m+1+k)}
y^{\b(k-1)}(\yb_{\ad(m)}\star\yb^{\bd(m+1+k)})\right)\nn\\
&&\sim T\left(\l^{\a\b} v_{\a\b(k)\bd(k+1)}
y^{\b(k-1)}\yb^{\bd(k+1)}\right)\ =\ 0\ .\eea
Finally, the $\l^{\c\dd}$-contribution to $[\e',\widehat B]_\pi$
reads
\bea&& {b_1\over 4} \int_0^1 dt\int_0^1 t'dt' \sum_{m,n}(-t')^m
i^{m+n}f_m f_n v^{\a(m)\ad(m)} v^{\b(n)\bd(n)}\times\nn\\&&
\times~ \l^{\c\dd}\left\{y_\c
\yb_{\dd}\,,\,z^\a\left((z_{\a(m+1)}\yb_{\ad(m)}e^{it'yz})\star
(y_{\b(n)}\yb_{\bd(n)})\right)_{Z\ra tZ}\right\}_\star\ ,\eea
which we evaluate at $Z=0$ using
\be \l^{\c\dd}\{y_\c\yb_{\dd},z^\a\widehat f\}_\star|_{Z=0}\ =\
-2i\l^{\a\dd}\yb_{\dd}\widehat f\ ,\ee
resulting in contributions to $T[\e',\widehat B]_\pi|_{Z=0}$ of
the form
\bea&& T\left(\l^{\a\dd}\yb_{\dd} v^{\a(m)\ad(m)}
v_{\a(m+1)\b(k)\bd(m+1+k)}
y^{\b(k)}(\yb_{\ad(m)}\star\yb^{\bd(m+1+k)})\right)\nn\\
&&\sim T\left(\l^{\a\dd}\yb_{\dd} v_{\a\b(k)\bd(k+1)}
y^{\b(k)}\yb^{\bd(k+1)}\right)\ =\ 0\ ,\eea
which concludes the proof of \eq{Ipr2}.

Interestingly enough, we have found a stronger result, namely that
\be C'\ =\ C'(V)\ ,\qquad V\ =\ y v\yb\ ,\ee
preserves $SO(3,2)$ to second order in the curvature expansion for
general initial conditions $C'(V)$. We expect $SO(3,2)$ to be
broken down to some $\mathbf g_r$ with $r\leq 6$ at third order.


\scs{Discussion}\label{sec:6}


We have seen that, while the field equations written in traditional
form in spacetime are highly complicated, and indeed not even
available explicitly at present, we can the nonetheless solve them
exactly by exploiting their relative simple form in terms of master
fields. Doing so, we have found the first exact solution of higher
spin gauge theory in $D>3$ other than the anti de Sitter spacetime.

The solution presented here forms a consistent background  for the
first-quantized topological open twistor string description of
higher spin gauge theory \cite{Johan}. It would be interesting to
study this $1+1$ dimensional field theory which provides a framework
for a dual twistor space interpretation of the solution.

It would also be useful to determine the holographic interpretation
of our solution. This requires, however, the knowledge of an
on-shell action and a systematic way to label the fluctuation fields
that takes into account the infinite dimensional nature of the
underlying symmetries. While progress has been made in that
direction \cite{Vasiliev:2005zu} much remains to be done to develop
fully the necessary tools to facilitate the fluctuation analysis.

Next, we note that while our solution is time dependent, its
cosmological interpretation is not straightforward. To begin with,
an understanding of the key concepts of horizons and singularities
are very much based on the geodesic equation of motion for test
particles, yet we do not know its higher spin covariant counterpart.

A key feature of the theory relevant in dealing with both
holographic and spacetime geometric aspects is the presence of
nonlocalities in the spacetime field equations. These nonlocalities
should be manageable, however, once we work with master fields. For
example, using this approach, we have already found certain
zero-form charges that provide a set of labels for classical
solutions. The master field approach should also be used to
construct a higher spin dressed version of a spacetime line element,
which in turn should be embedded into a manifestly higher spin
covariant infinite dimensional geometry.

Finally, it is interesting to compare our solution with the
$SO(3,1)$ invariant solution to the gauged ${\cal N}=8$, $D=4$
supergravity found in \cite{Hertog:2004rz}. To this end, we first
observe that the minimal bosonic models we have studied here are
consistent truncations of the higher-spin gauge theory based on
$shs(8|4)\supset osp(8|4)$ \cite{hs1} which contains, respectively,
$35_++35_-$ scalars and pseudo-scalars in the supergravity multiplet
and $1+\bar 1$ scalar and pseudo-scalar in an $s_{\rm max}=4$
multiplet that we refer to as the Konishi multiplet. The truncations
to the Type A and Type B models keep the Konishi scalar and
pseudo-scalar, respectively \cite{on124}. Thus we see that an
important difference between the solution of \cite{Hertog:2004rz}
and ours is that different scalars have been activated. The full
consequences of this difference remains to be investigated.

{\bf  Acknowledgements:}

We thank M. Nishimura for participation at an early stage of this
work, and we are grateful to M. Bianchi, J. Engquist, D. Francia, P.
Howe, R. Leigh, H. Lu, T. Petkou, C. Pope, P. Rajan, A. Sagnotti, B.
Sundborg and M. Vasiliev for discussions. We have benefitted from
the mutual hospitality of our home institutes. The work of E.S. is
supported in part by NSF Grant PHY-0314712 and the work of P.S. is
supported in part by INTAS Grant 03-51-6346.


\begin{appendix}



\scs{Conventions and Notation}\label{App:conv}

We use the conventions of \cite{hs1} in which the $SO(3,2)$
generators obey
\be [M_{AB},M_{CD}]\ =\ i\eta_{BC}M_{AD}+\mbox{$3$ more}\ ,\qquad
M_{AB}\ =\ (M_{AB})^\dagger\ ,\label{so32}\ee
with $\eta_{AB}={\rm diag}(--+++)$. The commutation relations
decompose into
\be [M_{ab},M_{cd}]_\star\ =\ 4i\eta_{[c|[b}M_{a]|d]}\ ,\qquad
[M_{ab},P_c]_\star\ =\ 2i\eta_{c[b}P_{a]}\ ,\qquad
[P_a,P_b]_\star\ =\ iM_{ab}\ .\label{so32b}\ee
The oscillator realization is taken to be
 \be
 M_{ab}\ =\ -\frac18 \left[~ \s_{ab})^{\a\b}y_\a y_\b+
 (\bar\s_{ab})^{\ad\bd}\yb_{\ad}\yb_{\bd}~\right]\ ,\qquad P_{a}\ =\
 \frac14 (\s_a)^{\a\bd}y_\a\yb_{\bd}\ ,\label{mab}
 \ee
where the van der Warden symbols obey
 \bea
  (\s^{a})_{\a}{}^{\ad}(\sb^{b})_{\ad}{}^{\b}&=& \y^{ab}\d_{\a}^{\b}\
 +\ (\s^{ab})_{\a}{}^{\b} \ ,\qquad
 (\sb^{a})_{\ad}{}^{\a}(\s^{b})_{\a}{}^{\bd}\ =\ \y^{ab}\d^{\bd}_{\ad}\
 +\ (\sb^{ab})_{\ad}{}^{\bd} \ ,\nn\w2
 \ft12 \e_{abcd}(\s^{cd})_{\a\b}&=& i(\s_{ab})_{\a\b}\ ,\qquad \ft12
 \e_{abcd}(\sb^{cd})_{\ad\bd}\ =\  -i(\sb_{ab})_{\ad\bd}\ ,\w2
 ((\s^a)_{\a\bd})^\dagger&=& (\bar\s^a)_{\ad\a}\ =\ (\s^a)_{\a\ad}\ ,\qquad
 ((\s^{ab})_{\a\b})^\dagger\ =\ (\bar\s^{ab})_{\ad\bd}\ ,
 \eea
with Minkowski space-time metric $\eta_{ab}={\rm diag}(-+++)$, and
spinor conventions $A^\a=\epsilon^{\a\b}A_\b$,
$A_\a=A^\b\epsilon_{\b\a}$, and $(A^\a)^\dagger=\bar A^{\ad}$.

The $SO(3,2)$-valued connection is expressed as
 \be
  \O\ =\ \frac1{4i}
 dx^\mu\left[\omega_\mu^{\a\b}~y_\a y_\b
 +{\bar\omega}_\mu{}^{\dot\a\dot\b}~{\bar y}_{\dot\a}{\bar y}_{\dot\b}
 + 2 e_\mu^{\a\dot\b}~y_\a {\bar y}_{\dot\b}\right]\
 .\label{Omega}
 \ee
The components of ${\cal R}=d\O+\O\wedge\star \O$ are
 \bea
 {\cal R}_{\a\b}&=& d\o_{\a\b} +\o_{\a\c}\wedge\o_{\b}{}^{\c}+
 e_{\a\dd}\wedge e_{\b}{}^{\dd}\ ,
 \label{rab}\w2
 {\cal R}_{\dot\a\dot\b}&=& d\bar{\o}_{\ad\bd}
 +\bar{\o}_{\ad\cd}\wedge\bar{\o}_{\bd}{}^{\cd}
 +e_{\d\ad}\wedge e^{\d}{}_{\bd}\ ,
 \label{radbd}\w2
 {\cal R}_{\a\dot\b}&=&  de_{\a\bd}+ \o_{\a\c}\wedge
 e^{\c}{}_{\bd}+\bar{\o}_{\bd\dd}\wedge e_{\a}{}^{\dd}\
 .\label{rabd}
 \eea
Defining
 \be
 \o^{\a\b}\ =\ -\ft14(\s_{ab})^{\a\b}~\o^{ab}\ ,
 \quad
 \o^{\dot\a\dot\b}\ =\ -\ft14({\bar\sigma}_{ab})^{\dot\a\dot\b}~\o^{ab}\ ,
 \quad
 e^{\a\dot\a}\ =\ -\ft{\lambda}2(\s_{a})^{\a \dot\a}~e^{a}\ ,
 \label{convert}
 \ee
where $\lambda$ is the inverse radius of the AdS$_4$ vacuum, and
converting the spinor indices of the curvatures in the same way,
gives
 \be
 {\cal R}^{ab}\ =\ d\o^{ab}+\o^a{}_c\o^{cb} +\lambda^2
 e^a\wedge e^b\ ,\qquad {\cal R}^a\ =\ d e^a+\o^a{}_b e^b\ .
 \label{curvcomp}
 \ee
It follows that the AdS$_4$ vacuum $\O_{(0)}$ is characterized by
 \be
 R_{(0)\m\n,\r\s}\ = \
 -\lambda^2 \left( g_{(0)\mu\rho}\bar g_{(0)\nu\sigma}-
 \bar g_{(0)\nu\rho}\bar g_{(0)\mu\sigma} \right)\ ,
 \qquad R_{(0)\mu\nu}\ =\ -3\lambda^2~g_{(0)\mu\nu}\label{ads}\ .
\ee
%


\scs{Gaussian Integration Formulae}


To compose coset representatives we make use of the formula
\be e^{iy a \yb}\star e^{iy b \yb}\ =\ \frac{1}{\det (1+\bar a b)}
\exp{ i \left[y(a{1\over 1+\bar b a}+b{1\over 1+\bar a b})\yb-y{a
\bar b\over 1+a\bar b}y+\yb {\bar b a\over 1+\bar b a}\yb\right]}\
,\label{gauss1}\ee
where we use matrix notation, \emph{e.g.} $y a\yb=y^\a
a_\a{}^{\bd}\yb_{\bd}$ and $y \bar a b y=y^\a \bar a_\a{}^{\bd}
b_{\bd}{}^\c y_\c$, with $a_{\a\bd}=a^\mu(\s_\mu)_{\a\bd}$ and
$\bar a_{\ad\b}=a^\mu (\s_\mu)_{\ad\b}$. In deriving this formula,
we use \eq{star} and perform the integrals over holomorphic and
anti-holomorphic variables \emph{independently}. In doing so, the
exponentials remain separately linear in integration variables,
allowing the use of the identity
\be \int {d^2\x d^2\eta\over (2\pi)^2)} e^{i\eta^\a
\left[\x_\a+u_\a(\bar\xi,\bar\eta)\right]+i\xi^\a
v_\a(\bar\xi,\bar\eta))}\ =\ e^{iu^\a(\bar\xi,\bar\eta)
v_\a(\bar\xi,\bar\eta)}\ ,\label{gauss1b}\ee
and its analog for anti-holomorphic variables. Other useful
formulae are
\bea e^{iy a\yb}\star e^{i\r y}\star e^{iya\yb}&=& {1\over
(1+a^2)^2}\exp i\left[{2ya\yb+(1-a^2)\r y\over 1+a^2}\right]\ ,\label{gauss4}\\
(e^{-itza\yb}e^{ityz})\star e^{iya \yb}&=& {1\over (1-ta^2)^2}
\exp i\left[{(1-a^2)t y z+ (1-t)ya\yb\over 1-ta^2}\right]\
.\label{gauss7} \eea
The relation \eq{zz} follows from the lemma
\bea &&e^{i\left[ty z +\r  y +\t z\right]}\star e^{i\left[t'y z
+\r'  y +\t'  z\right]}\label{gauss2}\\
&=& e^{i\left[(t+t'-2tt')y z+
\big((1-t')\r+(1-t)\r'+t\t'-t'\t\big) y
+\big((1-t')\t+(1-t)\t'+t\r'-t'\r\big) z+(\r+\t)
(\r'-\t')\right]}\ .\nn\eea
Finally, to evaluate $K_\m$ on the solution, we need
\bea &&\left.e^{i y a \yb}\star e^{i(sy z+\r y+\tau
z)}\star e^{-iya\yb}\right|_{Z=0}\nn\\[5pt]
&=& {1\over (1-(1-2s)a^2)^2} ~\exp i\left[ {\r
((1+a^2)y+2a\yb)-2a^2\r \t\over 1-(1-2s)a^2}\right]\
,\label{gauss3}\eea
where $a^2=a^\m a_\m$.


\scs{Second-Order Corrections to the Field Equations}\label{app:c}


The second-order source terms $\widehat P^{(2)}_\m$ and $\widehat
J^{(2)}_{\m\n}$ in the field equations are given by
\cite{Us:analysis}
\bea P_\m^{(2)} &=& \Phi\star\bar{\pi}(W_\m)- W_\m\star \Phi
\nn\w2 && +\left.\Bigg(\Phi\star\bar{\pi}({\widehat e}_\m{}^{(1)})
-\widehat{e}_\m^{(1)}\star \Phi+\widehat{\Phi}^{(2)}\star
\bar{\pi}(e_\m)-e_\m\star\widehat{\Phi}^{(2)}\Bigg)\right|_{Z=0}\
, \la{p2}\w4 J_{\m\n}^{(2)}&=& -\Bigg[
\left.\Bigg(\widehat{e}_\m^{(1)}\star \widehat{e}_\n^{(1)}+
\{e_\m,\widehat{e}_\n^{(2)}\}_{\star} +\{e_\m,
\widehat{W}_\n^{(1)}\}_{\star}+ \{W_\m,
\widehat{e}_\n^{(1)}\}_{\star} \Bigg)\right|_{Z=0} \nn\w2 &&
+\left.\Bigg(i
R_{\m\n}{}^{\a\b}\widehat{A}^{(1)}_\a\star\widehat{A}^{(1)}_\b
+\mbox{h.c.} \Bigg)\right|_ {Z=0} +W_\m\star W_\n \Bigg] -(\m
\leftrightarrow \n)\ , \la{j2} \eea
where the hatted quantities are defined as\footnote{In eq. \eq{a2}
the last term was inadvertently omitted in \cite{Us:analysis}.}
\bea
\hA^{(1)}_\a  &=&   -{i b_1\over 2} z_\a \int_0^1 t dt~
\F(-tz,\yb)\kappa(tz,y)\ , \la{a1}\w4
\widehat{A}^{(2)}_\a &=& z_\a \int_0^1 tdt\left(
\widehat{A}^{(1)\b}\star \widehat{A}^{(1)}_\b\right)_{Z\ra tZ}
+\zb^{\bd}\int_0^1 t dt \left[
\widehat{A}^{(1)}_{\a},\widehat{A}^{(1)}_{\bd}\right]_{\star Z\ra
t Z}\nn\w3 && -{ib_1\over 2} \int_0^1 t dt \left(\widehat
\Phi^{(2)}\star\kappa\right)_{Z\ra tZ}\ ,\label{a2} \w4
{\widehat\Phi}^{(2)}&=& \int_0^1dt\Bigg(z^\a\left(\Phi\star
{\bar\pi}(\hA_\a^{(1)}) - \hA_\a^{(1)}\star \Phi \right)_{Z\ra tZ}
+{\zb}^{\ad} \left(\Phi\star {\pi}({\widehat A}_{\ad}^{(1)}) -
\hA_{\ad}^{(1)}\star \Phi \right)_{Z\ra tZ} \Bigg)\ ,\qquad
\label{Phi2}\w4
\widehat{e}^{(1)}_\mu &=& -i e_\m^{\a\ad} \int_0^1 {dt\over t}
\Bigg( \left[{\yb}_{\ad},\hA_\a^{(1)}\right]_*
+\left[\hA_{\ad}^{(1)},y_\a\right]_* \Bigg)_{Z\ra tZ}\ ,
\label{emu1}\w4
\widehat{W}^{(1)}_\mu &=& -i\int_0^1{dt\over t}
\Bigg(\left[{\partial W_\m\over \partial
y^\a},\hA^{\a(1)}\right]_* +\left[\hA^{\ad (1)} ,{\partial
W_\m\over \partial \yb^{\ad}}\right]_* \Bigg)_{Z\ra tZ}\
,\label{w1}\\\w4
\widehat{e}^{(2)}_\mu&=&-i e_\m^{\a\ad} \int_0^1 {dt\over t}
\Bigg( \left[{\yb}_{\ad},\hA_\a^{(2)}\right]_* +\left[{\widehat
A}_{\ad}^{(2)},y_\a\right]_* \Bigg)_{Z\ra tZ}\label{emu2} \w3
&& -e_\m^{\a\ad} \int_0^1 {dt\over t} \int_0^1 {dt'\over t'}
\Bigg( \hA^{\b(1)}\star \Big(\frac{\partial}{\partial
z^\b}-{\partial\over\partial y^\b}\Big) \Big(
\left[{\yb}^{\phantom{()}}_{\ad},\hA_\a^{(1)}\right]_*
                     +\left[\hA_{\ad}^{(1)},y^{\phantom{()}}_\a\right]_*
                     \Big)_{Z\ra t'Z}
\nn\w3
&& \qquad\qquad\qquad\qquad + \hA^{\bd(1)}\star
\Big({\partial\over\partial \zb^{\bd}}+{\partial\over\partial
\yb^{\bd}}\Big) \left(
\left[{\yb}^{\phantom{()}}_{\ad},\hA_\a^{(1)}\right]_*
+\left[\hA_{\ad}^{(1)},y^{\phantom{()}}_\a\right]_*
                     \right)_{Z\ra t'Z}
\nn\w3
&& \qquad\qquad\quad + \Big({\partial\over\partial
z^\b}+{\partial\over\partial y^\b}\Big)\left(
\left[{\yb}^{\phantom{()}}_{\ad},\hA_\a^{(1)}\right]_*
                     +\left[\hA_{\ad}^{(1)},y^{\phantom{()}}_\a\right]_*
                     \right)_{Z\ra t'Z} \star \hA^{\b (1)}
\nn\w3
&& \qquad\qquad\quad + \Big({\partial\over\partial
\zb^{\bd}}-{\partial\over\partial \yb^{\bd}}\Big)\left(
\left[{\yb}^{\phantom{()}}_{\ad},\hA_\a^{(1)}\right]_*
                     +\left[\hA_{\ad}^{(1)},y^{\phantom{()}}_\a\right]_*
                     \right)_{Z\ra t'Z} \star \hA^{\bd(1)}
                     \Bigg)_{Z\ra tZ}\ , \nn\eea
where the replacement $(z,\zb)\ra(tz,t\zb)$ in a quantity must be
performed {\it after} the quantity has been written in Weyl
ordered form i.e. after the $\star$ products defining the quantity
has been performed.


\scs{Analysis of the $Q$-Function}\label{app:Q}


Expanding the denominator of \eq{Q}, and splitting into even and
odd parts in $a^2$, as
\be Q\ =\ Q_++Q_-\ ,\qquad Q_\pm(-a^2)\ =\ \pm Q_\pm(a^2)\ ,\ee
we find
\bea Q_+&=& {-(1-a^2)^2}\sum_{p=0}^\infty {-4\choose
2p}a^{4p}\left[\int_0^1dt~
t^{2p}\left(q^{(+)}(t)-tq^{(-)}(t)\right)\right]^2\
,\\[5pt]
Q_-&=& {(1-a^2)^2}\sum_{p=0}^\infty {-4\choose
2p+1}a^{4p+2}\left[\int_0^1dt~
t^{2p+1}\left(tq^{(+)}(t)-q^{(-)}(t)\right)\right]^2\ ,\eea
with $q^{(\pm)}(t)$ given by \eq{qplus} and \eq{qminus}.
Performing the integrals we arrive at
\bea Q_+&=&-{(1-a^2)^2\over 4}\sum_{p=0}^\infty {-4\choose
2p}a^{4p}\left(\sqrt{1-{\nu\over 2p+1}}-\sqrt{1+{\nu\over
2p+3}}\right)^2\label{Qpl}\\[8pt]
Q_-&=&{(1-a^2)^2\over 4}\sum_{p=0}^\infty {-4\choose
2p+1}a^{4p+2}\left(\sqrt{1-{\nu\over 2p+3}}-\sqrt{1+{\nu\over
2p+3}}\right)^2\ ,\label{Omi}\eea
that are valid for
\be -3\leq{\rm Re}~\nu\leq 1\ .\label{rangle}\ee
Expanding in $\nu$,
\be Q_+\ =\ \sum_{n=2}^\infty \nu^n Q_{+,n}\ ,\qquad Q_-\ =\
\sum_{n=2}^\infty \nu^{2n} Q_{+,2n}\ ,\ee
the first non-trivial contribution is found to be
\bea Q_{+,2}&=&-{(1-a^2)^2\over 8} \left({d^2\over da^4}
a^4\right)^2 \left[{}_3 F_2(1,1,4;3,3;-a^2)+{}_3
F_2(1,1,4;3,3;a^2)\right]\ ,\\[8pt]
Q_{-,2}&=&{(1-a^2)^2\over 8} \left[{}_3 F_2(2,2,4;3,3;-a^2)-{}_3
F_2(2,2,4;3,3;a^2)\right]\ .\eea
We note that parity acts in tangent space by exchanging
holomorphic and anti-holomorphic oscillators and, moreover, by
exchanging $\widehat \Phi$ with $\widehat\Phi$ or $-\widehat\Phi$
in the Type A and Type B models, respectively. Since the spin
$s=2,4,\dots$ Weyl tensors $C_{\a_1\dots\a_{2s}}$ can be taken to
be even under parity, this means that the Type B model is
invariant under $\phi\rightarrow-\phi$, or, equivalently, under
$\nu\rightarrow-\nu$, while this need not be, and is indeed not, a
symmetry in the Type A model.


\end{appendix}



\newpage
\begin{thebibliography}{99}



\bibitem{vasiliev} M.A. Vasiliev, ``{Consistent equations for
interacting gauge fields of all spins in $3+1$ dimensions}'',
Phys. Lett. {\bf 243} (1990) 378.

\bibitem{Vas:star} M.~A.~Vasiliev, ``Higher spin gauge theories:
star-product and AdS space'', {\tt hep-th/9910096}.

\bibitem{Vas:2003} M.A. Vasiliev, ``{ Nonlinear equations for symmetric massless
higher spin fields in $(A)dS_d$}'', Phys. Lett. B {\bf 567} (2003)
139, {\tt hep-th/0304049}.

\bibitem{Sagnotti:2005ns}
A.~Sagnotti, E.~Sezgin and P.~Sundell, ``On higher spins with a
strong Sp(2,R) condition'', {\tt hep-th/0501156}.

\bibitem{Bekaert:2005vh}
X.~Bekaert, S.~Cnockaert, C.~Iazeolla and M.~A.~Vasiliev,
``Nonlinear higher spin theories in various dimensions'', {\tt
hep-th/0503128}.

\bibitem{Us:analysis} E. Sezgin and  P. Sundell, ``{Analysis of higher spin
field equations in four dimensions}'', JHEP {\bf 0207} (2002) 055,
{\tt hep-th/0205132}.

\bibitem{Sezgin:2003pt}
E.~Sezgin and P.~Sundell, ``Holography in 4D (super) higher spin
theories and a test via cubic  scalar couplings'', JHEP {\bf 0507}
(2005) 044, {\tt hep-th/0305040}.

\bibitem{Kristiansson:2003xx}
  F.~Kristiansson and P.~Rajan,
  ``Scalar field corrections to AdS(4) gravity from higher spin gauge
  theory,''
  JHEP {\bf 0304} (2003) 009,
  {\tt hep-th/0303202}.

\bibitem{Vas:more} M.~A.~Vasiliev, ``More on equations of motion for interacting massless fields
of all spins in 3+1 dimensions'', Phys.\ Lett.\ B {\bf 285} (1992)
225.

\bibitem{5d} E. Sezgin and  P. Sundell, ``{Doubletons and 5D higher
spin gauge theory}'',  JHEP {\bf 0109} (2001) 036, {\tt
hep-th/0105001}.

\bibitem{5dn4} E. Sezgin and  P. Sundell, ``{Towards
massless higher spin extension of D=5, N=8 gauged supergravity}'',
JHEP {\bf 0109} (2001) 025, {\tt hep-th/0107186}.

\bibitem{7d} E. Sezgin
and P. Sundell, ``{7D bosonic higher spin theory: symmetry algebra
and linearized constraints}'', Nucl. Phys. B {\bf 644} (2002) 303
[Erratum-ibid. B {\bf 660} (2003) 403], {\tt hep-th/0112100}.

\bibitem{Vasiliev:2004cm}
  M.~A.~Vasiliev,
  ``Higher spin superalgebras in any dimension and their representations,''
  JHEP {\bf 0412} (2004) 046,
  {\tt hep-th/0404124}.

\bibitem{Cederwall:2004cf}
  M.~Cederwall,
  ``AdS twistors for higher spin theory,''
  AIP Conf.\ Proc.\  {\bf 767} (2005) 96,
  {\tt hep-th/0412222}.


\bibitem{Bars:2005ze}
  I.~Bars and M.~Picon,
  ``Single twistor description of massless, massive, AdS, and other interacting
  particles,''
  {\tt hep-th/0512091}.

\bibitem{Bars:2001uy}
  I.~Bars and S.~J.~Rey,
  ``Noncommutative Sp(2,R) gauge theories: A field theory approach to  two-time
  physics,''
  Phys.\ Rev.\ D {\bf 64} (2001) 046005,
  {\tt hep-th/0104135}.

\bibitem{Johan} J. Engquist and P. Sundell, ``Brane partons and
singleton strings'', {\tt hep-th/0508124}.


\bibitem{sundborg} B. Sundborg, ``Stringy gravity, interacting tensionless
massless higherspins'', Nucl. Phys. Proc. Suppl. {\bf 102} (2001)
113, {\tt hep-th/0103247}.

\bibitem{sundell2} E. Sezgin and P. Sundell, ``Massless higher spins and holography'',
Nucl. Phys. {\bf B644} (2003) 303; Erratum-ibid. B {\bf 660}
(2003) 403, {\tt hep-th/0205131}.

\bibitem{polyakovklebanov} I.R. Klebanov and A.M. Polyakov, ``AdS dual of the
critical O(N) vector model'', Phys. Lett. B {\bf 550} (2002) 213,
{\tt hep-th/0210114}.

\bibitem{Sagnotti:2003qa}
A.~Sagnotti and M.~Tsulaia, ``On higher spins and the tensionless
limit of string theory'', Nucl.\ Phys.\ B {\bf 682} (2004) 83,
{\tt hep-th/0311257}.

\bibitem{Beisert:2004di}
N.~Beisert, M.~Bianchi, J.~F.~Morales and H.~Samtleben, ``Higher
spin symmetry and N = 4 SYM'', JHEP {\bf 0407} (2004) 058, {\tt
hep-th/0405057}.

\bibitem{Prokushkin:1998bq}
S.~F.~Prokushkin and M.~A.~Vasiliev, ``Higher-spin gauge
interactions for massive matter fields in 3D AdS space-time'',
Nucl.\ Phys.\ B {\bf 545} (1999) 385, {\tt hep-th/9806236}.

\bibitem{Bolotin:1999fa}
K.~I.~Bolotin and M.~A.~Vasiliev, ``Star-product and massless free
field dynamics in AdS(4)'', Phys.\ Lett.\ B {\bf 479} (2000) 421,
{\tt hep-th/0001031}.


\bibitem{Varna} E.~Sezgin and P.~Sundell, ``On an Exact Cosmological
Solution of Higher Spin Gauge Theory'', {\tt hep-th/}.


\bibitem{vas:supertrace} M.~A.~Vasiliev, ``Extended higher spin
superalgebras and their realizations in terms of quantum
operators'', Fortsch. Phys. {\bf 36} (1988) 33.

\bibitem{Vasiliev:2005zu}
  M.~A.~Vasiliev,
  ``Actions, charges and off-shell fields in the unfolded dynamics approach,''
  {\tt hep-th/0504090}.

\bibitem{Hertog:2004rz}
T.~Hertog and G.~T.~Horowitz, ``Towards a big crunch dual,'' JHEP
{\bf 0407} (2004) 073, {\tt hep-th/0406134}; ``Holographic
description of AdS cosmologies,''
  JHEP {\bf 0504}, (2005) 005,
  {\tt hep-th/0503071}.

\bibitem{hs1} E. Sezgin and P. Sundell, ``Higher spin $N=8$
supergravity'', JHEP {bf 9811} (1998) 016, {\tt hep-th/9805125}.

\bibitem{on124} J.~Engquist, E.~Sezgin and P.~Sundell,
  ``On N = 1,2,4 higher spin gauge theories in four dimensions,''
  Class.\ Quant.\ Grav.\  {\bf 19} (2002) 6175,
  {\tt hep-th/0207101}.

\end{thebibliography}
\end{document}